\documentclass[preprint,12pt]{aastex}

\usepackage{amsmath}

\newcommand{\beq}{\begin{equation}}           
\newcommand{\eeq}{\end{equation}}             
\newcommand{\bfi}{\begin{figure}}           
\newcommand{\efi}{\end{figure}}             

\shortauthors{Dan\v{e}k \& Heyrovsk\'y}

\shorttitle{Analysis of $n$-point-mass Lenses}

\begin{document}

\title{Image-plane Analysis of $n$-point-mass Lens Critical Curves and Caustics}
	
\author{Kamil Dan\v{e}k and David Heyrovsk\'y}

\affil{Institute of Theoretical Physics, Faculty of Mathematics and Physics, \mbox{Charles University in Prague}, Czech Republic; \email{kamil.danek@utf.mff.cuni.cz}\email{heyrovsky@utf.mff.cuni.cz}}

\begin{abstract}

The interpretation of gravitational microlensing events caused by planetary systems or multiple stars is based on the $n$-point-mass lens model. The first planets detected by microlensing were well described by the two-point-mass model of a star with one planet. By the end of 2014, four events involving three-point-mass lenses had been announced. Two of the lenses were stars with two planetary companions each; two were binary stars with a planet orbiting one component. While the two-point-mass model is well understood, the same cannot be said for lenses with three or more components. Even the range of possible critical-curve topologies and caustic geometries of the three-point-mass lens remains unknown. In this paper we provide new tools for mapping the critical-curve topology and caustic cusp number in the parameter space of $n$-point-mass lenses. We perform our analysis in the image plane of the lens. We show that all contours of the Jacobian are critical curves of re-scaled versions of the lens configuration. Utilizing this property further, we introduce the cusp curve to identify cusp-image positions on all contours simultaneously. In order to track cusp-number changes in caustic metamorphoses, we define the morph curve, which pinpoints the positions of metamorphosis-point images along the cusp curve. We demonstrate the usage of both curves on simple two- and three-point-mass lens examples. For the three simplest caustic metamorphoses we illustrate the local structure of the image and source planes.

\end{abstract}

\keywords{gravitational lensing: micro --- planetary systems --- stars: multiple}

\section{Introduction}
\label{sec:intro}

The effect of the gravitational field of an astrophysical object on the passage of light from a background source was first computed for a point-mass-like object \citep{einstein36,link36,link37}, which provided the simplest model of a gravitational lens. The model can be used in some cases as a first approximation for describing the strong lensing of a quasar by a galaxy. Its main application, however, has been in Galactic gravitational microlensing, where it accurately describes the lensing of a background star by another star passing in the foreground \citep[e.g.,][]{paczynski96}.

The more than ten thousand observed microlensing events include several hundred cases of lensing by binary stars \citep[lists can be found in][]{alcock_etal00,jaroszynski02,jaroszynski_etal04,jaroszynski_etal06,jaroszynski_etal10,skowron_etal07}. In addition, there are more than 30 cases of microlensing by a star with a planet \citep[e.g.,][]{bond_etal04,beaulieu_etal06}. Planets are currently the primary goal of microlensing surveys, which are sensitive to Solar-system analogues including Earth-mass planets in AU-scale orbits. Binary stars and stars with a planet can be both described by two-point-mass microlensing, which has been well understood theoretically for the past twenty years.

\cite{chang_refsdal84} first described an extension of the single point-mass lens obtained by adding a constant external shear. Such a system has extended caustics with different geometries in the weak and strong shear regimes. \cite{schneider_weiss86} performed a detailed analysis of a two-point-mass lens with equal masses. They demonstrated that there were three distinct regimes for the critical curve and caustic: close, intermediate, and wide (in order of increasing separation of binary components). Transitions between neighboring regimes involved beak-to-beak metamorphoses of the caustic. \cite{erdl_schneider93} studied the lensing by two independent objects generally in different lens planes. As a special case they included the analysis of the single-plane (thin) two-point-mass lens with an arbitrary mass ratio, showing that the regimes and metamorphoses were the same as in the equal-mass case. This result was corroborated by \cite{witt_petters93}. From the range of subsequent papers we mention here the analysis by \cite{dominik99} of special cases of the binary lens.

However, already at $n=3$ observations overtook theory. Three-component lensing was detected first by \cite{gaudi_etal08} and second by \cite{han_etal13}. Both cases involved microlensing by a star with two planets. Most recently, \cite{gould_etal14} and \cite{poleski_etal14} documented two events with microlensing by a binary star with a planet orbiting one of the components. Despite these observational advances, we still have no general insight into the possible regimes, types of critical curves and caustics permitted by the three-point-mass lens model. Nevertheless, a number of theoretical papers have been published on different aspects of triple-lens microlensing.

The first step beyond two-point-mass lensing was taken by \cite{grieger_etal89}, who studied the effect of an additional constant shear and demonstrated the appearance of the swallow-tail and butterfly caustic metamorphoses. This particular model was studied later also by \cite{witt_petters93}. The general triple-lens equation was discussed by \cite{rhie02}. Other works concentrated on specific types of triple lenses. Most frequent among them are papers studying lensing by a star with two planets \citep[e.g.,][]{gaudi_etal98,han_etal01,ryu_etal11,song_etal14,zhu_etal14}. Several papers explored lensing by a binary star with a single planet \citep[e.g.,][]{lee_etal08,han08a}. Finally, the possibility of detecting exo-moons in microlensing events involving a star with a planet with a moon was discussed by \cite{han_han02}, \cite{han08b}, and \cite{liebig_wambsganss10}.

Much of the literature on general $n$-point-mass lenses revolved around the question of the maximum possible number of images generated by such a lens. Individual contributions include \cite{rhie97}, \cite{mao_etal97}, \cite{rhie01,rhie03}, \cite{khavinson_neumann06}, \cite{asada09b}, \cite{luce_etal14}, and \cite{sete_etal15}. Pivotal among these are \cite{rhie03}, who conjectured the maximum number of images was $5\,(n-1)$, and \cite{khavinson_neumann06}, who proved the conjecture.

Other papers used perturbative methods to study the $n$-point-mass lens formed by a star with $n-1$ planets. \cite{bozza99} explored the caustics of such a system, and \cite{asada09a} presented perturbative solutions of the lens equation. Finally, for the general $n$-point-mass lens \cite{bozza00a,bozza00b} studied the geometry of caustics in the close and wide limits.

The initial goal of the research behind the present paper was to systematically explore the properties of three-point-mass lenses. Following the methods used by \cite{erdl_schneider93} and \cite{witt_petters93} for the general two-point-mass lens, we set out to map critical-curve topologies and caustic geometries in the parameter space of the three-point-mass lens. Some early results appeared in \cite{danek10}, \cite{danek_heyrovsky11}, and \cite{danek_heyrovsky14}. In the course of our work we devised an efficient method for tracking the changes in critical-curve topology and caustic cusp number in the image plane of an arbitrary $n$-point-mass lens. We present these results here, while their application to the analysis of simple three-point-mass lens models appears separately in a companion paper \citep{danek_heyrovsky15}.

We start in \S~\ref{sec:n-point} by describing the studied lens model, a thin $n$-point-mass gravitational lens with no external shear and no convergence due to continuous matter. We transform the lens equation to complex formalism following \citep{witt90}, which translates the analysis of lens properties to the study of polynomials, curves, and real and complex functions in the complex plane.

Our approach is based on investigating the properties of the Jacobian of the lens, viewed in \S~\ref{sec:Jacobian} as a surface (real function) over the complex image plane. The main new insight is the correspondence of Jacobian contours with critical curves of zoomed-in or zoomed-out lens configurations, as described in \S~\ref{sec:Jacobian_correspondence}. The nature of the critical curve, and the changes of its topology in the parameter space of the lens are discussed in \S~\ref{sec:critical}.

In \S~\ref{sec:caustic} we study the caustic of the lens, paying particular attention to the changes in cusp number during caustic metamorphoses. We introduce two further tools for finding important points along Jacobian contours in the image plane: the cusp curve in \S~\ref{sec:cusp_curve} for tracking the positions of cusp images, and the morph curve in \S~\ref{sec:morph} for pinpointing the images of caustic-metamorphosis points. In \S~\ref{sec:butterfly} we give specific conditions for distinguishing butterfly and swallow-tail metamorphoses.

We demonstrate the introduced method in \S~\ref{sec:usage} with the example of a binary lens, and two triple lenses with swallow-tail and butterfly metamorphoses, respectively. Finally, we illustrate the local structures in the image and source planes in the vicinity of the three simplest caustic metamorphoses of the $n$-point-mass lens. The main results are summarized in \S~\ref{sec:summary}.

\section{Lensing by an $n$-point-mass lens}
\label{sec:n-point}

In this work we study a composite gravitational lens formed by an association of $n$ objects, which we approximate by point masses with masses $M_j,\,j=1 \ldots n$. We retain the usual thin-lens approximation, in which variations in the line-of-sight distance of individual objects from the observer are neglected. The entire lens is thus assumed to lie at a single distance $D_{l}$ from the observer and $D_{ls}$ from the background source, so that \mbox{$D_{s}=D_{l}+D_{ls}$} is the distance from the observer to the source. The angular positions {\boldmath $\theta_j$} of the individual objects (lens components) in the plane of the sky, the position $\boldsymbol\theta$ of an image formed by the lens, and the position $\boldsymbol\beta$ of the source are all defined relative to a specific point of the lens. The usual choices for this point are: the position of a lens component, the center of mass of the lens, or its geometric center. While the equations in this work are independent of the choice, in the presented examples we use the geometric center as the origin. All object positions must be distinct, $\boldsymbol\theta_j\neq\boldsymbol\theta_k$ for all $j\neq k$, in order to qualify as an $n$-point-mass lens. In the thin-lens approximation any pair of objects exactly aligned with the observer acts as a single object with the total mass of the pair, reducing the number of lens components by one.

The lens defines a characteristic scale in the plane of the sky: the angular Einstein radius corresponding to the total mass of the lens. It is given by
\beq
\theta_E=\sqrt{\frac{4\,G\,M_{tot}}{c^2}\frac{D_{ls}}{D_{l}\,D_{s}}}\,,
\label{eq:Einstein}
\eeq
where $G$ is the gravitational constant, $c$ is the speed of light, and \mbox{$M_{tot}=\sum_{j=1}^{n}{M_j}\,$} the total mass of the lens. In the absence of external shear and convergence due to continuous matter, the relation between the source and image positions is expressed by the lens equation
\beq
\boldsymbol\beta=\boldsymbol\theta-\theta_E^2\sum^{n}_{j=1} {\mu_j\,\frac{\boldsymbol\theta-\boldsymbol\theta_j}{|\boldsymbol\theta-\boldsymbol\theta_j|^2}}\,.
\label{eq:lenseq-n}
\eeq
Here $\mu_j\equiv M_j/M_{tot}$ are the fractional masses in units of total mass (hence $\sum^{n}_{j=1}\mu _j=1 $).

In order to study the properties of the lens it is advantageous to switch from vectorial to complex notation \citep{witt90}, in which all angular positions in the plane of the sky are expressed in terms of complex variables. If we additionally measure all angular positions in units of $\theta_E$, the positions of the individual components are \mbox{$z_j\equiv(\theta_{jI}+{\rm i}\,\theta_{jII})/\theta_E$}, the image position \mbox{$z\equiv(\theta_I+{\rm i}\,\theta_{II})/\theta_E$}, and the source position \mbox{$\zeta\equiv(\beta_I+{\rm i}\,\beta_{II})/\theta_E$}, where we used Roman indices to mark vector components, and ${\rm i}$ is the imaginary unit. The inverse transformation from complex to vectorial notation is obtained by combining a position with its complex conjugate. For example, the image position \mbox{$\boldsymbol\theta=(z+\bar{z},\,-{\rm i}\,z+{\rm i}\,\bar{z})\;\theta_E/2$}, where $\bar{z}$ denotes the complex conjugate of $z$. In matrix notation the transformation $\boldsymbol\theta\to z$ can be described by $(z,\bar{z})=C\,(\theta_I,\theta_{II})$ with
\beq
C=\frac{1}{\theta_E}\left(
\begin{array}{rr}
1 & {\rm i} \\
1 & {-\rm i} \\
\end{array}
\right)\,.
\label{eq:matrix}
\eeq
The same matrix transforms similarly {\boldmath $\theta_j$} $\to z_j$ and $\boldsymbol\beta\to \zeta$. The opposite transformation $z\to\boldsymbol\theta$ can be described using the inverse matrix $(\theta_I,\theta_{II})=C^{-1}\,(z,\bar{z})$, given explicitly by
\beq
C^{-1}=\frac{\theta_E}{2}\left(
\begin{array}{rr}
1 & 1 \\
{-\rm i} & {\rm i} \\
\end{array}
\right)\,.
\label{eq:inverse_matrix}
\eeq
Matrix $C^{-1}$ transforms also $z_j\to$ {\boldmath $\theta_j$} and $\zeta\to\boldsymbol\beta$ in an analogous way\footnote{In case the positions are already in units of $\theta_E$, set $\theta_E=1$ in equations~(\ref{eq:matrix}) and (\ref{eq:inverse_matrix}).}. This description will be useful for studying the Jacobian of the lens mapping and its properties in \S~\ref{sec:Jacobian}.

The lens equation~(\ref{eq:lenseq-n}) attains a simpler form in complex formalism,
\beq
\zeta=z-\sum^{n}_{j=1}{\frac{\mu_j }{\bar{z}-\bar{z}_j}}\,.
\label{eq:lenseq-n_complex}
\eeq
The positions of images for a given source position are usually found by transforming equation~(\ref{eq:lenseq-n_complex}) to polynomial form purely in terms of $z$. This can be achieved by taking the complex conjugate of equation~(\ref{eq:lenseq-n_complex}), expressing $\bar{z}$ in terms of $z$ and $\zeta$ from it, and substituting back into equation~(\ref{eq:lenseq-n_complex}), obtaining
\beq
\zeta=z-\sum^{n}_{j=1}{\frac{\mu_j }{\bar{\zeta}-\bar{z}_j+\sum^{n}_{k=1}{\frac{\mu_k }{z-z_k}}}}\;.
\label{eq:lenseq-n_pre-polynomial}
\eeq
The expression on the right-hand side can be converted to a rational function of $z$ by gradual transformation to a common denominator. Multiplying by the final denominator leads to a polynomial equation of degree $n^2+1$ for the image position $z$ \citep[e.g.,][]{witt90}.

Such an equation always yields a set of $n^2+1$ solutions. However, not all of these are true images. Additional solutions were introduced by the substitution for $\bar{z}$ when deriving equation~(\ref{eq:lenseq-n_pre-polynomial}), which is not an equivalent step. Only those solutions that satisfy the original lens equation~(\ref{eq:lenseq-n_complex}) correspond to images of the lensed source. The actual number of images for $n>1$ ranges from $n+1$ to $5\,n-5$ in increments of two \citep{rhie03, khavinson_neumann06}.

The number of images depends on the position of the source with respect to the caustic, which is a characteristic curve of the lens in the source plane \citep[e.g.,][]{schneider_weiss86}. When a point source enters the caustic, a pair of images appears and branches off from a point on the critical curve of the lens in the image plane. Conversely, when a point source exits the caustic, a pair of images meets and vanishes at the critical curve. In special cases, when the point source enters or exits the caustic at an intersection point of two or more sections of the caustic, two or more pairs of images may appear or vanish simultaneously at different points along the critical curve.

From the description above, it follows that the critical curve $z_{cc}$ is a degenerate image of the caustic $\zeta_c$. This property is usually used in reverse to find the caustic of the lens by setting $z=z_{cc}$ in equation~(\ref{eq:lenseq-n_complex}). In mathematical terms, the critical curve is defined as the zero set of the Jacobian determinant corresponding to the lens equation~(\ref{eq:lenseq-n_complex}).

For a given image of a point source its flux amplification is given by the reciprocal absolute value of the Jacobian determinant, and its parity is given by the sign of the determinant. Since the Jacobian determinant plays a pivotal role in the analysis of gravitational lensing, we explore its properties in more detail in the following section.

\section{The Jacobian and its properties}
\label{sec:Jacobian}

\subsection{Definition and global character}
\label{sec:Jacobian_definition}

The Jacobi matrix in real vectorial lensing notation is defined as
\beq
J_0\equiv\frac{\partial\left(\beta_I\,,\beta_{II}\right)}{\partial \left(\theta_I\,,\theta_{II}\right)}\,.
\label{eq:Jacobian_real}
\eeq
The usual definition of the Jacobi matrix in complex notation is
\beq
J\equiv\frac{\partial\left(\zeta\,,\bar{\zeta}\right)}{\partial \left(z\,,\bar{z}\right)}\,,
\label{eq:Jacobian_complex}
\eeq
where $z$ and $\bar{z}$ are taken as independent variables \citep{witt90}. The two matrices are connected by the simple transformation
\beq
J=C\,J_0\,C^{-1}\,,
\label{eq:Jacobian_transformation}
\eeq
with the matrices $C$ and $C^{-1}$ defined in equations~(\ref{eq:matrix}) and (\ref{eq:inverse_matrix}), respectively. Since equation~(\ref{eq:Jacobian_transformation}) conserves the determinant, ${\rm det}\,J={\rm det}\,J_0\,$, all properties of ${\rm det}\,J_0 (\boldsymbol\theta)$ as a real function over the real plane are identical to those of ${\rm det}\,J\,(z)$ as a real function over the complex plane. The determinant (hereafter the ``Jacobian") can thus be viewed and explored as a simple two-dimensional surface.

In the case of our studied lens equation~(\ref{eq:lenseq-n_complex}), the Jacobian simplifies to
\beq
{\rm det}\,J\,(z)=\partial_z\zeta\, \partial_{\bar{z}}\bar{\zeta} - \partial_{\bar{z}}\zeta\, \partial_z\bar{\zeta} = 1-\left|\partial_z\bar{\zeta}\right|^2=1-\left|\,\sum^{n}_{j=1}{\frac{\mu _j }{(z-z_j)^2}}\,\right|^{\,2}\,,
\label{eq:Jacobian}
\eeq
where we used abbreviated notation for partial derivatives, e.g., $\partial_z\zeta\equiv{\partial \zeta}/{\partial z}$ \citep{witt90,witt_petters93}. From the final expression we immediately see that ${\rm det}\,J\,(z)\leq 1$ over the entire complex plane. The Jacobian is a continuous smooth function everywhere except at $n$ poles at the positions of the lens components, where ${\rm det}\,J\,(z_j)=-\infty$. Far from all the lenses, for $|z-z_j|\gg 1$, it approaches unity, ${\rm det}\,J\,(z)\to 1$. The set of points with zero Jacobian forms the critical curve of the lens, ${\rm det}\,J\,(z_{cc})=0$.

\subsection{Stationary points: maxima and saddles}
\label{sec:stationary_points}

The stationary points of the Jacobian fulfill the condition $\partial_z\,{\rm det}\,J=0$. Since the Jacobian is a real function, $\partial_{\bar{z}}\,{\rm det}\,J=0$ is then implied automatically. Using equations~(\ref{eq:Jacobian}) and (\ref{eq:lenseq-n_complex}) we obtain the expression
\beq
\partial_z\,{\rm det}\,J=-\partial_{\bar{z}}\zeta\,\partial^2_z\bar{\zeta}=0\,,
\label{eq:Jacobian_stationary}
\eeq
where we used abbreviated notation for the second derivative, $\partial^2_z\bar{\zeta}\equiv{\partial^2 \bar{\zeta}}/{\partial z^2}$. Equation~(\ref{eq:Jacobian_stationary}) yields two types of solutions, $\partial_{\bar{z}}\zeta=0$ and $\partial^2_z\bar{\zeta}=0$. The nature of these solutions can be found from the Hessian matrix of the second derivatives along the real and imaginary axes, with components $h_{ij}=\partial^2 {\rm det}\,J_0/{\partial\theta_i\partial\theta_j}$ where $i, j=I, II$. Expressed in terms of complex derivatives, the Hessian matrix
\beq
H=C^T
  \left(
  \begin{array}{rr}
      \partial^2_z\,{\rm det}\,J & \partial^2_{z,\bar{z}}\,{\rm det}\,J \\
      \partial^2_{\bar{z},z}\,{\rm det}\,J & \partial^2_{\bar{z}}\,{\rm det}\,J\\
  \end{array}
  \right)
  C\,,
\label{eq:Hessian_matrix}
\eeq
where $C^T$ is the transpose of matrix $C$. Substituting for the Jacobian from equation~(\ref{eq:Jacobian}), we obtain the Hessian determinant
\beq
{\rm det}\,H=\frac{4}{\theta^4_E}
\left(|\partial^2_{\bar{z}}\zeta|^4- |\partial_{\bar{z}}\zeta|^2 |\partial^3_{\bar{z}}\zeta|^2\right)\,.
\label{eq:Hessian}
\eeq
Note that the Hessian has the opposite sign when compared with the determinant of the central matrix on the right-hand side of equation~(\ref{eq:Hessian_matrix}).

The stationary points with $\partial_{\bar{z}}\zeta=0$ have ${\rm det}\,H=4\,\theta^{-4}_E|\partial^2_{\bar{z}}\zeta|^4\geq 0$, hence they are local extrema if $\partial^2_{\bar{z}}\zeta\neq 0$. From equation~(\ref{eq:Jacobian}) we see that all points with $\partial_{\bar{z}}\zeta=0$ have ${\rm det}\,J=1$. Therefore, these points are always maxima of the Jacobian irrespective of the second derivative, achieving the same peak value as for $z\to\infty$. Note that the Jacobian value also implies that these maxima never lie on the critical curve. In terms of the lens parameters, the condition for Jacobian maxima is
\beq
\partial_z\bar{\zeta}=\sum^{n}_{j=1}{\frac{\mu_j }{(z-z_j)^2}}=0\,,
\label{eq:Maxima-npoint}
\eeq
where we used the complex conjugate of equation~(\ref{eq:lenseq-n_complex}). Multiplication by all denominators converts equation~(\ref{eq:Maxima-npoint}) to a polynomial of degree $2\,n-2$ with highest-degree-term coefficient equal to 1. The $n$-point-mass lens Jacobian thus has at most $2\,n-2$ different local maxima, since the polynomial generally may have multiple roots, for which $\partial^2_{\bar{z}}\zeta=\partial^2_z\bar{\zeta}=0$. In the limit when all lens components are mutually much closer than the Einstein radius, $|z_j|\ll1$, the Jacobian maxima lie at the centers of the small loops of the critical curve.

The stationary points with $\partial^2_z\bar{\zeta}=0$ have ${\rm det}\,H=-4\,\theta^{-4}_E|\partial_{\bar{z}}\zeta|^2 |\partial^3_{\bar{z}}\zeta|^2\leq 0$, hence they are saddle points \citep{schneider_weiss86,erdl_schneider93} unless $\partial_{\bar{z}}\zeta=0$ or $\partial^3_{\bar{z}}\zeta=0$. In the former case they are higher-order maxima, as shown above. In the latter case they are higher-order saddle points, such as the monkey saddle if $\partial^4_{\bar{z}}\zeta\neq0$. In terms of the lens parameters the condition for Jacobian saddle points can be written as
\beq
-\frac{1}{2}\,\partial^2_z\bar{\zeta}=\sum^{n}_{j=1}{\frac{\mu_j }{(z-z_j)^3}}=0\,,
\label{eq:saddle-npoint}
\eeq
where we used equation~(\ref{eq:lenseq-n_complex}) and complex conjugation. Multiplication by all denominators converts equation~(\ref{eq:saddle-npoint}) to a polynomial of degree $3\,n-3$ with highest-degree-term coefficient equal to 1. The $n$-point-mass lens Jacobian thus has at most $3\,n-3$ different saddle points. Some solutions of equation~(\ref{eq:saddle-npoint}) may correspond to higher-order maxima, or several solutions may correspond to a higher-order saddle. The saddle points play a key role in determining changes in the topology of the critical curve, as discussed further in \S~\ref{sec:critical}.

\subsection{Jacobian contours and critical curves}
\label{sec:Jacobian_correspondence}

The $n$-point-mass-lens Jacobian has a remarkable property we have not found mentioned in the literature on gravitational lensing. We demonstrate here that all level curves of the Jacobian, seen as contours in the image plane, correspond to critical curves of re-scaled lens configurations. We can show this by taking any level curve of the Jacobian surface, e.g., ${\rm det}\,J\,(z)=\lambda$, where the Jacobian range restricts the parameter $\lambda$ to real values $\lambda\leq 1$. Using the last expression for the Jacobian from equation~(\ref{eq:Jacobian}), the equation for the corresponding contour in the image plane is
\beq
1-\left|\,\sum^{n}_{i=1}{\frac{\mu_j }{(z-z_j)^2}}\,\right|^{\,2}=\lambda\,.
\label{eq:level_curve}
\eeq
If we subtract $\lambda$ and divide by $1-\lambda\,$, we get
\beq
1-\frac{1}{1-\lambda}\left|\,\sum^{n}_{j=1}{\frac{\mu_j }{(z-z_j)^2}}\,\right|^{\,2}=0\,.
\label{eq:Jacobian_correspondence}
\eeq
This step can be taken anywhere except exactly at maxima. There $\lambda=1$ and the corresponding contours are point-like. Now if we re-scale the image plane including the lens positions by a real positive factor,
\beq
z'=z\,\sqrt[4]{1-\lambda}\,,\;\;z_j'=z_j\,\sqrt[4]{1-\lambda}\,,
\label{eq:rescaling}
\eeq
equation~(\ref{eq:Jacobian_correspondence}) simplifies to
\beq
1-\left|\,\sum^{n}_{j=1}{\frac{\mu_j }{(z'-z'_j)^2}}\,\right|^{\,2}=0\,,
\label{eq:rescaled_critical}
\eeq
which is the critical-curve equation for a lens with components at $z'_j$.

One can proceed similarly the other way around: if we change the scale of a lens configuration by shifting all lens components from $z_j$ to $\alpha\,z_j$, where $\alpha$ is a real positive factor, we find that the critical curve is equal to the Jacobian contour ${\rm det}\,J=1-\alpha^4$ of the original configuration re-scaled by $\alpha$.

This property is especially useful for studies of the lensing behavior, and it further enhances the significance of the Jacobian surface. For any particular $n$-point-mass lens the surface includes as its contours not only the critical curve of the given configuration, but also the critical curves of all directly similar configurations, arbitrarily shrunk or expanded. For values $\lambda\to -\infty$ in equation~(\ref{eq:level_curve}) we get the critical curve of a set of $n$ individual widely separated lenses (the ``wide" limit), while for $\lambda\to 1^-$ we get the critical curve of a combined lens with all components approaching a single point (the ``close" limit), as can be seen from equation~(\ref{eq:rescaling}).

In Figure~\ref{fig:Jacobian_correspondence} we demonstrate the Jacobian contour $\leftrightarrow$ critical curve correspondence on the example of a two-point-mass lens (hereafter for simplicity ``binary'' lens) with masses $\mu_1=4/5$ and $\mu_2=1/5$ separated by $s_0=1$ Einstein radius. In the left panel we plotted the Jacobian surface ${\rm det}\,J\,(z)$, with its ``trunks" around the lens positions truncated at ${\rm det}\,J=-15$. On the surface we marked a sequence of Jacobian contours specified in the caption. The highest contours outside the upper self-intersecting ${\rm det}\,J=0.711$ contour each have two additional small loops around the positions of the two Jacobian maxima. The three saddle points of the Jacobian surface lie at the self-intersection points of this contour and of the lowermost ${\rm det}\,J=-11.0$ contour.

In the right panel we plotted the same contours as seen from above in the image plane, adding the innermost ${\rm det}\,J=-100$ curve that would lie below the plot in the left panel. For illustration, selected curves corresponding to Jacobian values marked in the caption in boldface are labeled by the re-scaled binary separation $s_0\,\sqrt[4]{1-{\rm det}\,J}$. The given contour is identically equal to the critical curve in a similarly re-scaled image plane.

The right panel may also be viewed as a sequence of critical-curve plots for lenses with the same mass ratio and different separations $s$, with the position in the image plane measured in units of $s$ instead of $\theta_E$. Finally, the sequence of curves also corresponds to holding the absolute angular separation $s\,\theta_E$ fixed and changing the angular Einstein radius $\theta_E$ (e.g., by varying the total mass of the lens $M_{tot}$, or any of the distances $D_s$, $D_l$, $D_{ls}$).

Note that the positions of Jacobian maxima and saddle points scale together with the lens positions, i.e., their relative configuration stays fixed with the lens components. This can be seen from equations~(\ref{eq:Maxima-npoint}) and (\ref{eq:saddle-npoint}), in which any scaling factor in the denominator drops out due to the zero right-hand side, so that both equations retain their form. The correspondence with contours of the Jacobian surface in the left panel of Figure~\ref{fig:Jacobian_correspondence} provides the best illustration: the maxima and saddle points are defined globally just like the lens positions, irrespective of which of the contours is the critical curve.

\section{The $n$-point-mass lens critical curve}
\label{sec:critical}

\subsection{Critical-curve topology}
\label{sec:critical_structure}

The critical curve $z_{cc}$ is the set of all points in the image plane with zero Jacobian. Using the second expression for the Jacobian from equation~(\ref{eq:Jacobian}), we can write the general critical-curve condition in the compact form
\beq
\partial_{\bar{z}}\zeta\,\partial_z\bar{\zeta}=1\,.
\label{eq:critical-compact}
\eeq
If we multiply the last expression in equation~(\ref{eq:Jacobian}) by all the denominators, we see that  for $n\geq2$ the critical curve is a real algebraic curve of order $4\,n$. The same expression also indicates that the zero-Jacobian condition is satisfied if the derivative $\partial_z\bar{\zeta}$ lies anywhere along the unit circle. Hence, the points of the critical curve are the solutions of
\beq
\sum^{n}_{j=1}{\frac{\mu_j }{(z-z_j)^2}}=e^{-2\,{\rm i}\,\phi}\, ,
\label{eq:critical-npoint}
\eeq
where $\phi\in[0,\pi)$ is a phase parameter \citep{witt90}. The normalization of the phase is arbitrary; in our choice one gets the single-lens critical curve in the form $z_{cc}=z_1\pm\sqrt{\mu_1}\,e^{{\rm i}\,\phi}$.

The critical-curve equation~(\ref{eq:critical-npoint}) can be multiplied by all the denominators to yield a polynomial equation of degree $2\,n$. It follows that for any value of $\phi$ there are up to $2\,n$ different solutions of equation~(\ref{eq:critical-npoint}). By varying the phase from 0 to $\pi$ the solutions trace out $2\,n$ continuous critical-curve segments in the image plane. The choice of sign of the phase in equation~(\ref{eq:critical-npoint}) ensures that while tracing any segment in the sense of increasing $\phi$, the Jacobian is positive on its right side, negative on its left side. This can be shown by differentiating equations~(\ref{eq:Jacobian}) and (\ref{eq:critical-npoint}) to obtain
\beq
{\rm det}\,J\,(z_0\mp{\rm i\,d}z)=\pm4\,{\rm d}\phi\,,
\label{eq:critical_tracing}
\eeq
where $z_0$ is a point on the critical curve, and ${\rm d}z$ is a tangential shift along the critical curve corresponding to phase shift ${\rm d}\phi$. With this notation $z_0\mp{\rm i\,d}z$ are points offset perpendicularly to the right or left, respectively, from the tangential shift ${\rm d}z$.

The segments connect together in groups of 1 to $2\,n$ to form closed loops, the full set of which constitutes the critical curve. This follows from the continuous and periodic dependence of equation~(\ref{eq:critical-npoint}) on $\phi$, together with the fact that the critical curve avoids poles and complex infinity, $z\neq z_j,\infty$.

From the limitations on the number of segments per loop it follows that the number of loops of the critical curve may be as low as one (as in the case of the single lens or the intermediate binary lens), but not higher than $2\,n$ \citep{witt90}. Individual loops may lie outside one another (as in the wide binary lens), or some may lie nested in another (as in the close binary lens). The number of loops and their mutual position together define the topology of the critical curve.

The topology of the critical curve determines the structure of the caustic, which in turn determines the number of images as well as the structure of observable light curves in microlensing events. The topology depends on the parameters of the $n$-point-mass lens, i.e., on the masses and positions of the individual components. The actual number of relevant parameters for $n>1$ is reduced from $3\,n$ to $3\,n-4\,$, after accounting for the freedom of choice of the coordinate system (i.e., its origin, orientation, and position-measurement unit). Mapping the critical-curve topology in this parameter space is a crucial step toward understanding the nature of lensing by an $n$-point-mass lens.

\subsection{Identifying topology changes in parameter space}
\label{sec:topology_changes}

In order to map the critical-curve topology in the parameter space of the lens, we follow the approach of \cite{erdl_schneider93} and \cite{witt_petters93}, and identify parameter combinations that correspond to changes in the topology. These combinations define boundaries of regions with specific topologies in the parameter space. In order to find them we first note that individual loops of the critical curve are typically disconnected and have no self-intersections. Intersections occur only at the boundaries, where the topology changes by the splitting or merging of loops.

The critical curve, viewed as the ${\rm det}\,J=0$ Jacobian contour, separates regions of positive and negative Jacobian in the image plane. Therefore, at a simple intersection point on a critical curve where two loops meet, there is a facing pair of positive regions and a facing pair of negative regions. At the intersection the Jacobian peaks at zero when crossing from one negative region to the other. At the same time it drops there to zero when crossing from one positive region to the other. Hence, such an intersection must be a saddle point of ${\rm det}\,J$ \citep[e.g.,][]{schneider_weiss86}. In case three loops meet at an intersection point, the Jacobian does change sign but undergoes an inflexion when crossing it. The three-fold symmetry forms a monkey saddle on the surface, as mentioned in \S~\ref{sec:stationary_points}. Similarly, higher-order intersections of the critical curve correspond to higher-order saddles on the Jacobian surface. Generally, intersections of any order must simultaneously satisfy the saddle-point equation~(\ref{eq:saddle-npoint}) as well as the critical-curve equation~(\ref{eq:critical-npoint}).

The transitions in parameter space can be illustrated on the example of the binary lens in Figure~\ref{fig:Jacobian_correspondence}. For lens separations $s>1.861$ the critical curve consists of two separate loops around the lenses. At $s\approx1.861$ the critical curve passes through a saddle point, at which the two loops come into contact. For lens separations $1.861>s>0.733$ the critical curve consists of a single merged loop surrounding both lenses. At $s\approx0.733$ the critical curve self-intersects at a symmetric pair of saddle points, splitting off two small loops. For lens separations $s<0.733$ the critical curve consists of an outer loop plus two small loops surrounding the maxima of the Jacobian. Overall, for the given mass ratio the different topologies are separated by boundaries at $s\approx1.861$ and $s\approx0.733$, at which the critical curve passes through saddle points of the Jacobian.

As discussed in \S~\ref{sec:stationary_points}, the $n$-point-mass lens may have up to $3\,n-3$ different saddle points $z_{sadd}$. Each of them lies on a Jacobian contour ${\rm det}\,J\,(z)={\rm det}\,J\,(z_{sadd})$. While in the binary lens two of the three saddles lie on the same contour, for higher $n$ there may be in principle up to $3\,n-3$ different contours dividing the image plane into regions with different critical-curve topology. Taking into account the scaling properties discussed in \S~\ref{sec:Jacobian_correspondence}, this means that any lens configuration may undergo up to $3\,n-3$ topology changes when changing its scale from the ``wide" to the ``close" limit. By configuration we mean here fixed relative masses and positions of the components except an absolute scale. This property can greatly simplify the mapping of critical-curve topologies for different $n$-point-mass lens models, as demonstrated in the companion paper \citep{danek_heyrovsky15}.

\section{The $n$-point-mass lens caustic}
\label{sec:caustic}

\subsection{Caustic structure and cusp number}
\label{sec:caustic_structure}

The caustic $\zeta_c$ is a characteristic curve of the lens, defined as the set of all points in the source plane with infinite point-source amplification. Since the amplification of an image is the inverse absolute value of its Jacobian, the caustic must have an image with zero Jacobian. Indeed, in addition to regular images the caustic has a degenerate image, the critical curve $z_{cc}$.

If we know the critical curve, the caustic can be obtained simply from the lens equation
\beq
\zeta_c=z_{cc}-\sum^{n}_{j=1}{\frac{\mu_j }{\bar{z}_{cc}-\bar{z}_j}}\,.
\label{eq:caustic}
\eeq
Since the critical curve passes neither through lens positions, nor through complex infinity, the simple continuous transformation in equation~(\ref{eq:caustic}) preserves various properties of the critical curve. For example, each loop of the critical curve corresponds to a loop of the caustic, so that the number of loops is identical. The boundaries in parameter space corresponding to changes in critical-curve topology thus also correspond to changes in the number of caustic loops.

Unlike for single or binary lenses, the mapping between the critical curve and the caustic given by equation~(\ref{eq:caustic}) for lenses with $n\geq3$ components is not always one-to-one. Different points of the critical curve may be mapped onto the same point of the caustic. As a result, loops of the caustic may intersect each other, and individual loops may self-intersect. While following a path in parameter space, a caustic loop may move from the inside of another loop and cross it to the outside without a change in the number of loops. Hence, the mutual position of loops does not play the same role as in the critical curve, where the only contact between loops occurs at points where the loop number changes. The mutual position of caustic loops does play a role in identifying the number of images of the background source, which changes when crossing the caustic, as described in \S~\ref{sec:n-point}. The caustic thus plays the role of a boundary between regions of the source plane with different numbers of images.

Another distinction of the caustic from the critical curve is its lack of smoothness. Within our $n$-point-mass lens model given by equation~(\ref{eq:lenseq-n}), all loops of the caustic have cusps, sharp points at which the tangent-vector orientation changes by $\pi$. The smooth parts of the caustic between cusps are called folds. The number of cusps of a given caustic loop changes only under special local caustic metamorphoses.

The range of possible metamorphoses expands when increasing the number of lens components. The binary lens with $n=2$ permits only the simplest beak-to-beak metamorphosis, in which two tangent folds reconnect and form two facing cusps. This metamorphosis corresponds to a change in critical-curve topology; the boundaries are thus identical.

All higher-order metamorphoses occur without any accompanying change in the critical-curve topology. For lenses with $n=3$ components, two additional metamorphoses may occur. In the swallow-tail metamorphosis two cusps and a self-intersection arise from a point along a fold. In the butterfly metamorphosis two additional cusps and three self-intersections arise from a cusp.

In the rest of this section we concentrate on identifying the conditions under which caustic metamorphoses occur. In the parameter space they define boundaries that form a finer subdivision of the critical-curve topology regions discussed in \S~\ref{sec:topology_changes} according to the number of cusps of the caustic.

\subsection{Cusp condition}
\label{sec:cusps}

Cusps of the caustic are points at which the caustic tangent changes phase by $\pi$. Since the caustic is computed from the critical curve following equation~(\ref{eq:caustic}), we derive the tangent to the caustic from the tangent to the critical curve.

We start with the gradient of the Jacobian in complex notation,
\beq
G_z=\frac{\partial\,{\rm det}\,J}{\partial\,{\rm Re[}z{\rm ]}}+ {\rm i}\,\frac{\partial\,{\rm det}\,J}{\partial\,{\rm Im[}z{\rm ]}}=2\,\partial_{\bar{z}}\,{\rm det}\,J\,.
\label{eq:Jacobian_gradient}
\eeq
The critical curve is a contour of the Jacobian, hence its tangent is perpendicular to the gradient. We define the tangent to the critical curve as
\beq
T_z={\rm i}\,G_z=2\,{\rm i}\,\partial_{\bar{z}}\,{\rm det}\,J\,,
\label{eq:tangent_critical_0}
\eeq
evaluated along the critical curve \citep{witt90}. In this definition the absolute value of $T_z$ is equal to the absolute value of the gradient at the given point. This implies that the only points that may occur along the critical curve with $T_z=0$ are Jacobian saddle points. Critical-curve loops meet there for lens parameters at the boundaries studied in \S~\ref{sec:topology_changes}.

The orientation of $T_z$ is such that the Jacobian increases (i.e., is positive) to its right and decreases (i.e., is negative) to its left. For our lens equation~(\ref{eq:lenseq-n_complex}) we may use the complex conjugate of the left equality from equation~(\ref{eq:Jacobian_stationary}) to obtain
\beq
T_z=-2\,{\rm i}\,\partial_z\bar{\zeta}\,\partial^2_{\bar{z}}\zeta\,.
\label{eq:tangent_critical}
\eeq

Using the Jacobi matrix $J$ from equation~(\ref{eq:Jacobian_complex}) we compute the tangent to the caustic $T_\zeta$ from the transformation
\beq
\left(
\begin{array}{c}
T_\zeta \\
\bar{T}_\zeta \\
\end{array}
\right)
=J\,
\left(
\begin{array}{c}
T_z \\
\bar{T}_z \\
\end{array}
\right)
\,,
\label{eq:tangent_transformation}
\eeq
which gives us \citep{witt90}
\beq
T_\zeta=T_z\,\partial_z\zeta +\bar{T}_z\,\partial_{\bar{z}}\zeta\,.
\label{eq:tangent_caustic_0}
\eeq
Using lens equation~(\ref{eq:lenseq-n_complex}) and equation~(\ref{eq:tangent_critical}), we obtain
\beq
T_\zeta=2{\rm i}\,\left[(\partial_{\bar{z}}\zeta)^2\,\partial^2_z\bar{\zeta}- \partial_z\bar{\zeta}\,\partial^2_{\bar{z}}\zeta\right]\,.
\label{eq:tangent_caustic}
\eeq
As we trace the critical curve, the right-hand side of equation~(\ref{eq:tangent_caustic}) changes continuously, since the critical curve avoids the singularities at the lens positions. The only way in which the argument of $T_\zeta$ may change abruptly by $\pi$ is when $|T_\zeta|$ shrinks to zero. Therefore, cusps occur at points along the caustic where $T_\zeta=0$.

Using equation~(\ref{eq:tangent_caustic}) we conclude that in the image plane, cusps correspond to points along the critical curve which fulfill the cusp condition
\beq
(\partial_{\bar{z}}\zeta)^2\,\partial^2_z\bar{\zeta}= \partial_z\bar{\zeta}\,\partial^2_{\bar{z}}\zeta\,.
\label{eq:cusp_condition}
\eeq

\subsection{The cusp curve and its properties}
\label{sec:cusp_curve}

If we multiply equation~(\ref{eq:cusp_condition}) by $\partial^2_z\bar{\zeta}\,(\partial_z\bar{\zeta})^2$ and utilize the critical-curve equation~(\ref{eq:critical-compact}), we get
\beq
(\partial^2_z\bar{\zeta}\,)^2=|\,\partial^2_z\bar{\zeta}\,|^{\,2}\, (\partial_z\bar{\zeta}\,)^3\,.
\label{eq:cusp_condition_1}
\eeq
We note that for Jacobian saddle points $\partial^2_z\bar{\zeta}=0$, so that both sides of equations~(\ref{eq:cusp_condition}) and (\ref{eq:cusp_condition_1}) are equal to zero. If a saddle lies on the critical curve, it corresponds to cusp images. At such points in the image plane loops of the critical curve meet, while in the source  plane cusps meet in a beak-to-beak metamorphosis, as discussed in \S~\ref{sec:caustic_structure}.

For all other solutions we divide equation~(\ref{eq:cusp_condition_1}) by the first term on the right-hand side. From equation~(\ref{eq:critical-npoint}) we know that the last term on the right-hand side lies on the unit circle for critical-curve points. We may thus write the result as
\beq
\arg\left[\partial^2_z\bar{\zeta}\,\right]^2= \arg\left[\partial_z\bar{\zeta}\,\right]^3\,,
\label{eq:cusp_curve}
\eeq
an equation involving only the arguments (phases) of the powers of lens-equation derivatives. An equivalent formulation of equation~(\ref{eq:cusp_curve}) is
\beq
(\partial^2_z\bar{\zeta}\,)^2=4\,\Lambda\, (\partial_z\bar{\zeta}\,)^3\,,
\label{eq:cusp_curve_par}
\eeq
where $\Lambda$ is a real non-negative parameter, as seen from equation~(\ref{eq:cusp_condition_1}). In view of the one degree of freedom, equation~(\ref{eq:cusp_curve}) or (\ref{eq:cusp_curve_par}) describes a curve in the image plane which intersects the critical curve at the positions of cusp images. For brevity we call it the cusp curve.

If we compute the derivatives of $\bar{\zeta}$ using lens equation~(\ref{eq:lenseq-n_complex}), equations~(\ref{eq:cusp_curve}) and (\ref{eq:cusp_curve_par}) can be converted to
\beq
\arg\left[\sum^{n}_{j=1}{\frac{\mu_j }{(z-z_j)^3}}\right]^2=\arg\left[\sum^{n}_{j=1}{\frac{\mu_j }{(z-z_j)^2}}\right]^3\,,
\label{eq:cusp_curve1}
\eeq
and
\beq
\left[\sum^{n}_{j=1}{\frac{\mu_j }{(z-z_j)^3}}\right]^2=\Lambda\,\left[\sum^{n}_{j=1}{\frac{\mu_j }{(z-z_j)^2}}\right]^3\,.
\label{eq:cusp_curve_par1}
\eeq
The latter equation can be converted to a polynomial of degree $6\,n-6$, so that varying $\Lambda$ from 0 to $\infty$ yields up to $6\,n-6$ branches of the cusp curve. The cusp curve connects all important points of the image plane: the lens positions, Jacobian maxima, Jacobian saddle points, and complex infinity. The value $\Lambda=0$ corresponds uniquely to the saddle points (up to $3\,n-3$ different ones), and $\Lambda=\infty$ uniquely to the maxima (up to $2\,n-2$ different ones). Complex infinity corresponds to several roots at $\Lambda=1$. Similarly, each lens position $z_j$ corresponds to several roots at $\Lambda=1/\mu_j$.

Changing the scale in the image plane $(z,z_j)\to\alpha(z,z_j)$ leaves equations~(\ref{eq:cusp_curve1}) and (\ref{eq:cusp_curve_par1}) unchanged, which means the cusp curve is scale-invariant. Therefore, when added to a Jacobian-contour plot such as the right panel of Figure~\ref{fig:Jacobian_correspondence}, it will intersect all contours (i.e., re-scaled critical curves) at the positions of cusp images. A single plot can thus illustrate the number and distribution of cusps not only for the given lens configuration, but simultaneously also for all arbitrarily shrunk or expanded configurations. We point out here that the initial forms of the cusp condition in equations~(\ref{eq:cusp_condition}) and (\ref{eq:cusp_condition_1}) are not scale-invariant, hence they cannot be used in the same way.

For illustration we present in the left panel of Figure~\ref{fig:curves_binary} the cusp curve of the binary lens from Figure~\ref{fig:Jacobian_correspondence} with $\mu_1=4/5$ and $s_0=1$. The curve can be constructed from 6 branches starting at the saddle points and ending at the maxima, with some branches turning en route at the lens positions (two branches through each) or at complex infinity (two branches proceeding to either lens). The intersections of the cusp curve with the critical curve show the positions of cusp images, hence the corresponding caustic has six cusps.

Intersections with the other contours show us the positions of cusp images for re-scaled lens configurations (in this case for different binary separations $s$). The number of cusps changes in this case only through beak-to-beak metamorphoses that occur here for separations $s\approx1.861$ and $s\approx0.733$ (see also the discussion in \S~\ref{sec:topology_changes}). If we increase the separation to $s>1.861$, two additional cusps appear between the lenses, as shown by the contour closest to the lenses. The two loops of the critical curve each have four intersections, hence the two caustic loops have four cusps each. On the other hand, if we reduce the separation to $s<0.733$, two additional pairs of cusps appear outwards from the off-axis saddle points. The outer loop of the critical curve has four intersections (its caustic loop has four cusps); the two small loops around the maxima have three intersections each (their caustic loops have three cusps each).

The geometry of the cusp curve and the sequence of caustic transitions remains the same for any two-point-mass lens. Adding a third component to the lens opens up a range of different geometries, as illustrated by the examples in \S~\ref{sec:usage} and in \cite{danek_heyrovsky15}. The cusp curve then becomes a useful guide for understanding the occurrence of different caustic metamorphoses, and for keeping track of the cusp number on individual caustic loops.

\subsection{Caustic metamorphoses and the morph curve}
\label{sec:morph}

When changing the scale of a given lens configuration, the local appearance of two additional cusps other than at beak-to-beak metamorphoses may be detected in the image plane by finding points where the cusp curve is tangent to the local Jacobian contour. For a simple tangent point, contours adjacent from one side have no intersection with the cusp curve while contours adjacent from the opposite side each have two intersections. Such a configuration corresponds to the swallow-tail metamorphosis.

If the tangent point occurs at a simple self-intersection point of the cusp curve away from saddle points, contours adjacent from one side have one intersection with the cusp curve while contours adjacent from the opposite side each have three intersections. Such a configuration corresponds to the butterfly metamorphosis.

The appearance of additional cusps at a point on the caustic requires stationarity of the caustic tangent $T_\zeta$ at the corresponding point in the image plane when moving along the critical curve \citep[e.g.,][]{petters_etal01}. This requirement implies the condition
\beq
\left(T_z\,\partial_z+\bar{T}_z\,\partial_{\bar{z}}\right)\,T_\zeta=0\,,
\label{eq:swalbutt}
\eeq
which must be fulfilled in addition to the critical-curve equation~(\ref{eq:critical-compact}) and the cusp condition given by equation~(\ref{eq:cusp_condition}).

In the case of our lens equation~(\ref{eq:lenseq-n_complex}) we can substitute from equations~(\ref{eq:tangent_critical_0}) and (\ref{eq:tangent_caustic_0}) for the tangents $T_z$ and $T_\zeta$, respectively. Equation~(\ref{eq:swalbutt}) can then be written in the somewhat ungainly form
\beq
\partial_z\bar{\zeta}\,\partial^2_{\bar{z}}\zeta\left[\,\partial^2_{\bar{z}}\zeta\, \partial^2_z\bar{\zeta} - (\partial_{\bar{z}}\zeta)^2\,\partial^3_z\bar{\zeta}\,\right]= \partial_{\bar{z}}\zeta\,\partial^2_z\bar{\zeta}\left[\,\partial_z\bar{\zeta}\, \partial^3_{\bar{z}}\zeta -2\,\partial_{\bar{z}}\zeta\,\partial^2_{\bar{z}}\zeta\, \partial^2_z\bar{\zeta}\, \right]\,.
\label{eq:swalbutt-ugly}
\eeq
The terms before the brackets indicate that all stationary points of the Jacobian are among the solutions of equation~(\ref{eq:swalbutt}). Here we seek other solutions, such as the swallow-tail and butterfly metamorphosis points. In order to find them we simplify equation~(\ref{eq:swalbutt-ugly}) utilizing the critical-curve and cusp conditions.

We first multiply equation~(\ref{eq:swalbutt-ugly}) by $\partial^2_z\bar{\zeta}$ and substitute for $(\partial_{\bar{z}}\zeta)^2\,\partial^2_z\bar{\zeta}$ in the second terms on both sides of the equation using the cusp equation~(\ref{eq:cusp_condition}). Next we multiply the obtained equation by $\partial_{\bar{z}}\zeta$, which cancels out $\partial_z\bar{\zeta}$ on the left-hand side as well as in both terms on the right-hand side, due to the critical-curve equation~(\ref{eq:critical-compact}). Now we have
\beq
(\partial^2_{\bar{z}}\zeta)^2\,(\partial^2_z\bar{\zeta})^2 - \partial_z\bar{\zeta}\,(\partial^2_{\bar{z}}\zeta)^2\,\partial^3_z\bar{\zeta}= \partial_{\bar{z}}\zeta\,(\partial^2_z\bar{\zeta})^2\,\partial^3_{\bar{z}}\zeta- 2\,(\partial^2_{\bar{z}}\zeta)^2\,(\partial^2_z\bar{\zeta})^2\,.
\label{eq:swalbutt-semiugly}
\eeq
The second term on the right-hand side can be added to the first on the left-hand side. If we divide the equation by $|\,\partial^2_z\bar{\zeta}\,|^4$ and transfer the second term on the left-hand side to the right-hand side, we are left with the real equation
\beq
3={\rm Re\,[\,}2\,\partial_z\bar{\zeta}\,(\partial^2_z\bar{\zeta})^{-2}\, \partial^3_z\bar{\zeta}\,{\rm ]}\,.
\label{eq:morph_curve}
\eeq
Since the imaginary part of the expression in the square brackets is not constrained by the equation and thus may be arbitrary, we parameterize it as $3\,{\rm i}\,\Gamma$, where $\Gamma$ is a real parameter. We may now write
\beq
2\,\partial_z\bar{\zeta}\,\partial^3_z\bar{\zeta}= 3\,(\partial^2_z\bar{\zeta}\,)^2\,(1+{\rm i}\,\Gamma\,)\,.
\label{eq:morph_curve_par}
\eeq
We first note that in case the Jacobian has higher-order maxima or higher-order saddles (see \S~\ref{sec:stationary_points}), they will solve equation~(\ref{eq:morph_curve_par}) for any $\Gamma$. Maxima never lie on the critical curve, while saddles always correspond to topology-changing beak-to-beak metamorphoses (with more beaks involved in the case of higher-order saddles).

Setting these isolated points aside, equation~(\ref{eq:morph_curve}) is a real equation in the complex plane depending both on $z$ and $\bar{z}$, while equation~(\ref{eq:morph_curve_par}) is a complex parametric equation for $z$. Both describe the same curve, which we call the morph curve. It passes through the critical-curve points corresponding to all caustic-metamorphosis points, such as beak-to-beak, swallow-tail, and butterfly. In addition, it passes through the lens positions and complex infinity.

If we compute the derivatives of $\bar{\zeta}$ and $\zeta$ using lens equation~(\ref{eq:lenseq-n_complex}), equations~(\ref{eq:morph_curve}) and (\ref{eq:morph_curve_par}) can be converted to
\beq
1= {\rm Re}\,\left\{ \left[\,\sum^{n}_{j=1}{\frac{\mu_j}{(z-z_j)^2}}\right] \left[\,\sum^{n}_{j=1}{\frac{\mu_j}{(z-z_j)^3}}\right]^{-2} \left[\sum^{n}_{j=1}{\frac{\mu_j}{(z-z_j)^4}}\right]\right\}\,,
\label{eq:morph_curve1}
\eeq
and
\beq
\left[\,\sum^{n}_{j=1}{\frac{\mu_j}{(z-z_j)^2}}\right] \left[\sum^{n}_{j=1}{\frac{\mu_j}{(z-z_j)^4}}\right]= (1+{\rm i}\,\Gamma\,)\, \left[\,\sum^{n}_{j=1}{\frac{\mu_j}{(z-z_j)^3}}\right]^2\,,
\label{eq:morph_curve_par1}
\eeq
respectively. Multiplying the latter equation by all denominators converts it to a polynomial of degree $6\,n-6$, so that varying $\Gamma$ from $-\infty$ to $\infty$ yields up to $6\,n-6$ branches of the morph curve. The values $\Gamma=-\infty$ and $\Gamma=\infty$ correspond uniquely to the saddle points (up to $3\,n-3$ different ones). Each simple saddle point is crossed twice in perpendicular directions by the morph curve, monkey saddles are crossed symmetrically three times, higher-order saddles a corresponding higher number times. Turning around any saddle point, the arriving morph-curve parameters alternate between $\Gamma=-\infty$ and $\Gamma=\infty$. Complex infinity and the lens positions each correspond to a pair of roots at $\Gamma=0$.

From both versions of the morph-curve equation it is clear that changing the scale in the image plane $(z,z_j)\to\alpha(z,z_j)$ leaves equations~(\ref{eq:morph_curve1}) and (\ref{eq:morph_curve_par1}) unchanged, which means that even the morph curve is scale-invariant. We can therefore add it to Jacobian-contour plots and identify its intersections with the cusp curve away from the lens positions. Intersections at the saddle points identify the topology-changing beak-to-beak metamorphoses studied earlier. Simple intersections away from saddle points identify swallow-tail metamorphoses. Tangent intersections of a single morph-curve branch at a simple cusp-curve self-intersection point identify butterfly metamorphoses. For any of these points, the Jacobian contour passing through it identifies the scale of the lens configuration at which the metamorphosis occurs.

\subsection{Distinguishing butterflies from swallow tails}
\label{sec:butterfly}

In order to distinguish between swallow-tail and butterfly metamorphoses, we first check the condition for self-intersection points on the cusp curve. An arbitrarily small circle around such a point must have more than two intersections with the cusp curve defined by equation~(\ref{eq:cusp_curve1}). By expanding the equation we obtain the condition
\beq
3=2\,\partial_z\bar{\zeta}\,(\partial^2_z\bar{\zeta})^{-2}\, \partial^3_z\bar{\zeta}\,.
\label{eq:butterfly_self-intersection_condition}
\eeq
Comparing this self-intersection condition with the morph curve equation~(\ref{eq:morph_curve_par}), we find that all butterfly points must be morph-curve points with $\Gamma=0$. From the discussion above we know that for $\Gamma=0$ the morph curve has $2\,n$ roots at the lens positions plus 2 roots at complex infinity. It follows that a given lens configuration may have at most $4\,n-8$ butterfly points.

Butterfly metamorphoses additionally require the critical curve to be tangent to one of the branches of the cusp curve. Expanding the critical-curve equation~(\ref{eq:critical-compact}) to first order around its point $z_0$ away from saddles, we get its tangent
\beq
z_{cc}\simeq z_0+{\rm i}\,(\partial_{\bar{z}}\zeta\,\partial^2_z\bar{\zeta})^{-1}\,\varepsilon_1\,,
\label{eq:critical-expansion}
\eeq
where all derivatives are computed at $z_0$, and the real parameter $\varepsilon_1$ varies along the tangent. Similarly, we expand the cusp-curve equation~(\ref{eq:cusp_curve}) to first order around a self-intersection point away from maxima and lens positions. The tangents of its two branches are
\beq
z_{cusp_\pm}\simeq z_0+e^{\,{\rm i}(-2\psi\pm\pi)/4}\,\varepsilon_2\,,
\label{eq:cusp_curve_expansion}
\eeq
where the phase
\beq
\psi=\arg\left\{{\rm i}\left[3(\partial^2_z\bar{\zeta})^3-\partial^4_z\bar{\zeta}(\partial_z\bar{\zeta})^2\right] \left[(\partial_z\bar{\zeta})^2\partial^2_z\bar{\zeta}\,\right]^{-1}\right\}\,,
\eeq
all derivatives are computed at $z_0$, and the real parameter $\varepsilon_2$ varies along the tangent. For one of the two perpendicular tangents to be equal to the critical-curve tangent, the first-order terms of the expansions in equations~(\ref{eq:critical-expansion}) and (\ref{eq:cusp_curve_expansion}) must have the same argument in the complex plane (up to an integer multiple of $\pi$). Setting the arguments equal and using equations~(\ref{eq:critical-compact}) and (\ref{eq:cusp_condition_1}) we derive the condition
\beq
{\rm Im}\,[\,\partial^4_z\bar{\zeta}\,(\partial_z\bar{\zeta}\, \partial^2_z\bar{\zeta})^{-1}\,]\,=0.
\label{eq:butterfly_tangency}
\eeq
Any $\Gamma=0$ morph-curve intersection with the cusp curve has to be checked using this tangency condition, to make sure the metamorphosis is a butterfly rather than a pair of swallow tails.

In the rare case when the critical and cusp curves additionally have the same curvature at their tangent intersection point, the condition in equation~(\ref{eq:butterfly_tangency}) is not sufficient. The described technique then has to be extended by comparing higher-order expansions of the curves. Generally, a butterfly-like transition from one to three intersections occurs if the first difference between the expansions appears at an even order. Note that the equality of each order of the expansions introduces an additional constraint on the lens parameters. Such situations would therefore occur in lenses with a sufficiently high dimension of their parameter space (i.e., with a higher number of components).

Finally, we have to check whether we are at a simple self-intersection of the cusp curve, since equation~(\ref{eq:butterfly_self-intersection_condition}) is valid even for higher-order self-intersections with three or more intersecting branches. This final condition for the butterfly metamorphosis is
\beq
\partial^4_z\bar{\zeta}\neq3\,(\partial_z\bar{\zeta})^{-2}\,(\partial^2_z\bar{\zeta})^3\,.
\label{eq:simple_self-intersection_condition}
\eeq
Violation of this condition would lead to higher-order metamorphoses, which can be analyzed in a similar manner. Nevertheless, they do not occur in the triple lens and their exploration goes beyond the scope of this paper.

To summarize, the butterfly metamorphosis occurs when the critical curve intersects the cusp curve so that is passes as a tangent through a self-intersection point involving exactly two branches of the cusp curve. The conditions on the first through fourth derivatives of the lens equation are given by equations~(\ref{eq:critical-compact}), (\ref{eq:cusp_curve}), (\ref{eq:butterfly_self-intersection_condition}), (\ref{eq:butterfly_tangency}), and (\ref{eq:simple_self-intersection_condition}). Mapping the self-intersection back to the source plane using lens equation~(\ref{eq:lenseq-n_complex}) yields the position of the butterfly metamorphosis point on the caustic.

\section{Lens analysis using cusp and morph curves}
\label{sec:usage}

In the right panel of Figure~\ref{fig:curves_binary} we present the morph curve of the binary lens from Figure~\ref{fig:Jacobian_correspondence} plotted together with the cusp curve. Their five mutual intersection points include the two lens positions and the three Jacobian saddles. At the lens positions the incoming cusp-curve branches alternate with the morph-curve branches. At the (simple) saddle points the cusp curve crosses the intersecting morph-curve branches, identifying the occurrence of beak-to-beak metamorphoses for correspondingly re-scaled lens configurations. The morph curve has the same geometry for all two-point-mass lenses, marking the positions of the three beak-to-beak metamorphosis-point images along the respective Jacobian contours (re-scaled critical curves).

As mentioned above in \S~\ref{sec:caustic_structure}, swallow-tail and butterfly metamorphoses occur in $n$-point-mass lenses starting from $n=3$. We note here that they occur already for a two-point-mass lens with non-zero external shear, as shown by \cite{grieger_etal89}. Such a model is relevant to the present work as a limiting case of a two-point-mass lens weakly perturbed by a third component. In Figure~\ref{fig:curves_triple1} we illustrate the swallow-tail case as seen in the image plane of a triple equal-mass lens in a linear configuration with lens positions $\{z_1,z_2,z_3\}= \{-29/30,-2/30,31/30\}\,s$ for $s=0.5$. Here the scale parameter $s$ is one half of the separation of the two outer lenses.

The Jacobian surface has a generic triple-lens topology with 4 maxima and 6 saddle points, which correspond to beak-to-beak metamorphoses as indicated by the intersections of the cusp and morph curves in the right panel. In addition, there are 4 simple intersections of the cusp and morph curves in the right panel, seen also in the left panel as points where the cusp curve is tangent to Jacobian contours (two pairs above and below the axis between the maxima). Since they occur away from cusp-curve self-intersections, they correspond to swallow-tail metamorphoses.

We can now directly read off the sequence of gradual caustic changes from Figure~\ref{fig:curves_triple1}, as we vary the scale of the lens from the wide to the close limit. In the figure this corresponds to progressing from the lowest to the highest Jacobian contour, starting from the small loops around the lenses. In the wide limit the caustic has three loops with four cusps each. The loops corresponding to the two left bodies merge through a beak-to-beak metamorphosis forming a loop with six cusps. Another beak-to-beak metamorphosis merges it with the third-body loop, leaving a single 8-cusped loop.

The critical curve then reaches the symmetric pair of swallow-tail points closer to the lens axis. Two simultaneous swallow-tail metamorphoses add four cusps to form a single 12-cusped loop. Next comes the symmetric pair of saddle points near the two left maxima. Two simultaneous beak-to-beak metamorphoses split off two 3-cusped loops, leaving a third 10-cusped caustic loop. The symmetric pair of swallow-tail points further from the lens axis changes the third loop by two simultaneous swallow-tail metamorphoses removing four cusps, so that it is left with only six cusps. The symmetric pair of saddle points near the two right maxima leads to the final pair of simultaneous beak-to-beak metamorphoses, splitting off two more 3-cusped loops and thus removing two cusps from the 6-cusped loop. In the close limit the caustic has four 3-cusped loops and a single 4-cusped loop.

In Figure~\ref{fig:curves_triple2} we illustrate the butterfly metamorphosis in the image plane of a similar triple equal-mass lens, in a linear configuration but with symmetric lens positions $\{z_1,z_2,z_3\}= \{-1,0,1\}\,s$ for $s=0.5$. Even in this case the Jacobian has 4 maxima and 6 saddle points, which correspond to beak-to-beak metamorphoses indicated by the intersections of the cusp and morph curves in the right panel. In addition, there are 4 intersections along the imaginary axis in the right panel, each with a single branch of the morph curve passing through a self-intersection point of the cusp curve. In the left panel we see them as points where one branch of the cusp curve at a simple self-intersection point (away from lens positions and Jacobian maxima) is tangent to the local Jacobian contour. This confirms the intersections correspond to butterfly metamorphoses.

The sequence of gradual caustic changes is simpler here as we vary the scale of the lens from the wide to close limit, due to the higher degree of symmetry of the lens. In the wide limit the caustic has three loops around the lens components with four cusps each. All loops merge through two simultaneous beak-to-beak metamorphoses forming a loop with eight cusps. Two simultaneous butterfly metamorphoses due to the butterfly points closer to the real axis add four cusps to form a single 12-cusped loop. Four simultaneous beak-to-beak metamorphoses due to the four off-axis saddles then split off four 3-cusped loops, leaving a fifth 8-cusped caustic loop. Finally the outer pair of butterfly points causes two simultaneous butterfly metamorphoses that remove four cusps from the fifth loop, leading to the close limit with four 3-cusped loops and a single 4-cusped loop.

In Figure~\ref{fig:metamorphoses} we summarize the local behavior of the first three elementary caustic metamorphoses occurring in $n$-point-mass lenses \citep[see e.g.,][]{schneider_etal92,petters_etal01}. The left, central, and right pairs of columns illustrate the beak-to-beak (vicinity of central saddle point from Figure~\ref{fig:curves_binary}), the swallow-tail (swallow-tail point close to top right maximum from Figure~\ref{fig:curves_triple1}), and the butterfly (butterfly point just above central lens from Figure~\ref{fig:curves_triple2}) metamorphoses, respectively. The left panel of each pair shows the detail in the image plane including Jacobian contours with constant ${\rm det}\,J$ spacing. The right panel shows the corresponding detail of the caustic in the source plane. Rows correspond to different lens scales parameterized by $s$, as marked in each pair (top: before metamorphosis, middle: at metamorphosis, bottom: after metamorphosis with two new cusps). The image-plane details are plotted in units of $s$, so that the three critical curves in each column correspond to three different contours from the same Jacobian contour plot.

In addition to the three types of local behavior shown in Figure~\ref{fig:metamorphoses}, for a triple lens the only more complicated metamorphosis occurs at monkey saddles, where three loops of the critical curve meet. In the source plane they correspond to beak-to-beak metamorphoses with three ``beaks" meeting at a single point. An example can be found in the companion paper \citep{danek_heyrovsky15}.

In lenses with $n>3$ components higher-order saddles may occur, corresponding to a meeting of a higher number of ``beaks" in the source plane. In addition, higher-order metamorphoses may occur at cusp-curve self-intersections involving three or more cusp-curve branches. Their exploration, which may proceed in a similar manner as the butterfly analysis in \S~\ref{sec:butterfly}, is beyond the scope of this paper.

As a final caveat, note that all three scenarios illustrated in Figure~\ref{fig:metamorphoses} may be complicated in special situations. The beak-to-beak scenario may be altered if the cusp curve crosses the saddle as a tangent to the critical curve. While this would have no effect on the change in critical-curve topology, the caustic may possibly undergo no change in cusp number. In the swallow-tail and butterfly scenarios there may be no change in cusp number if the critical and cusp curves have the same curvature at the tangent point. The outcome of these situations can be studied by expanding both curves around the metamorphosis point and checking whether the first order at which the curves differ is even or odd.

All caustic metamorphoses studied here were demonstrated by changing the lens scale, which permits the usage of contour plots to visualize the full sequence. However, the metamorphoses occur even when changing arbitrary other parameters of the lens. Nevertheless, they happen exactly at the points identified by the presented analysis. All relevant equations and the sequence of caustic changes remain the same for any crossing of the respective parameter-space boundary.

\section{Summary}
\label{sec:summary}

In this work we studied $n$-point-mass lensing by exploring the geometry of the Jacobian, viewed as a two-dimensional surface defined over the image plane. If we plot the Jacobian contours in the image plane, the zero-Jacobian contour represents the critical curve of the lens. However, as we show in \S~\ref{sec:Jacobian_correspondence}, any contour of the Jacobian ${\rm det}\,J\,(z)=\lambda$ corresponds to the critical curve of a lens with component spacing re-scaled by the factor $\sqrt[4]{1-\lambda}$.

For a lens consisting of a given combination of point masses in a given formation, this property permits the simultaneous study of critical curves for all re-scaled configurations, from the single-lens close limit to the $n$-independent-lenses wide limit. A single contour plot thus illustrates the lens properties for a full 1D cut through the parameter space of the lens, instead of just for a single parameter combination.

We extend the approach further in \S~\ref{sec:caustic} by introducing two scale-invariant curves: the cusp curve given by equation~(\ref{eq:cusp_curve1}), and the morph curve given by equation~(\ref{eq:morph_curve1}). The cusp curve identifies the positions of cusp images along all the Jacobian contours. The morph curve pinpoints positions along the cusp curve, at which the number of cusps changes due to caustic metamorphoses.

For illustration, we present several simple applications in \S~\ref{sec:usage}. The overall structure of the curves for the binary lens shown in Figure~\ref{fig:curves_binary} is generic, valid for any mass ratio. However, adding a third component already opens up a broader range of possible structures and associated lens properties. We use triple-lens examples to illustrate two additional caustic metamorphoses, the swallow tail in Figure~\ref{fig:curves_triple1} and the butterfly in Figure~\ref{fig:curves_triple2}. Details of the image and source planes in Figure~\ref{fig:metamorphoses} demonstrate the characteristic features associated with these metamorphoses as well as the simple beak-to-beak metamorphosis.

In the companion paper \citep{danek_heyrovsky15} we use the presented approach for a systematic exploration of several simple triple-lens models. However, the concepts introduced here can be used also for more general studies. For example, a detailed analysis of the structure of critical, cusp, and morph curves would enable a full classification of local lensing behavior of triple, quadruple, etc. lenses. We conclude that image-plane analysis of Jacobian contour plots may prove to be a useful tool for studying and visualizing the nature of lensing by $n$-point-mass lenses.

\acknowledgements

We thank the anonymous referee for helpful suggestions. Work on this project was supported by Czech Science Foundation grant GACR P209-10-1318 and Charles University grant SVV-260089.

\clearpage
\bfi
\hspace{-2mm}
\includegraphics[scale=.285]{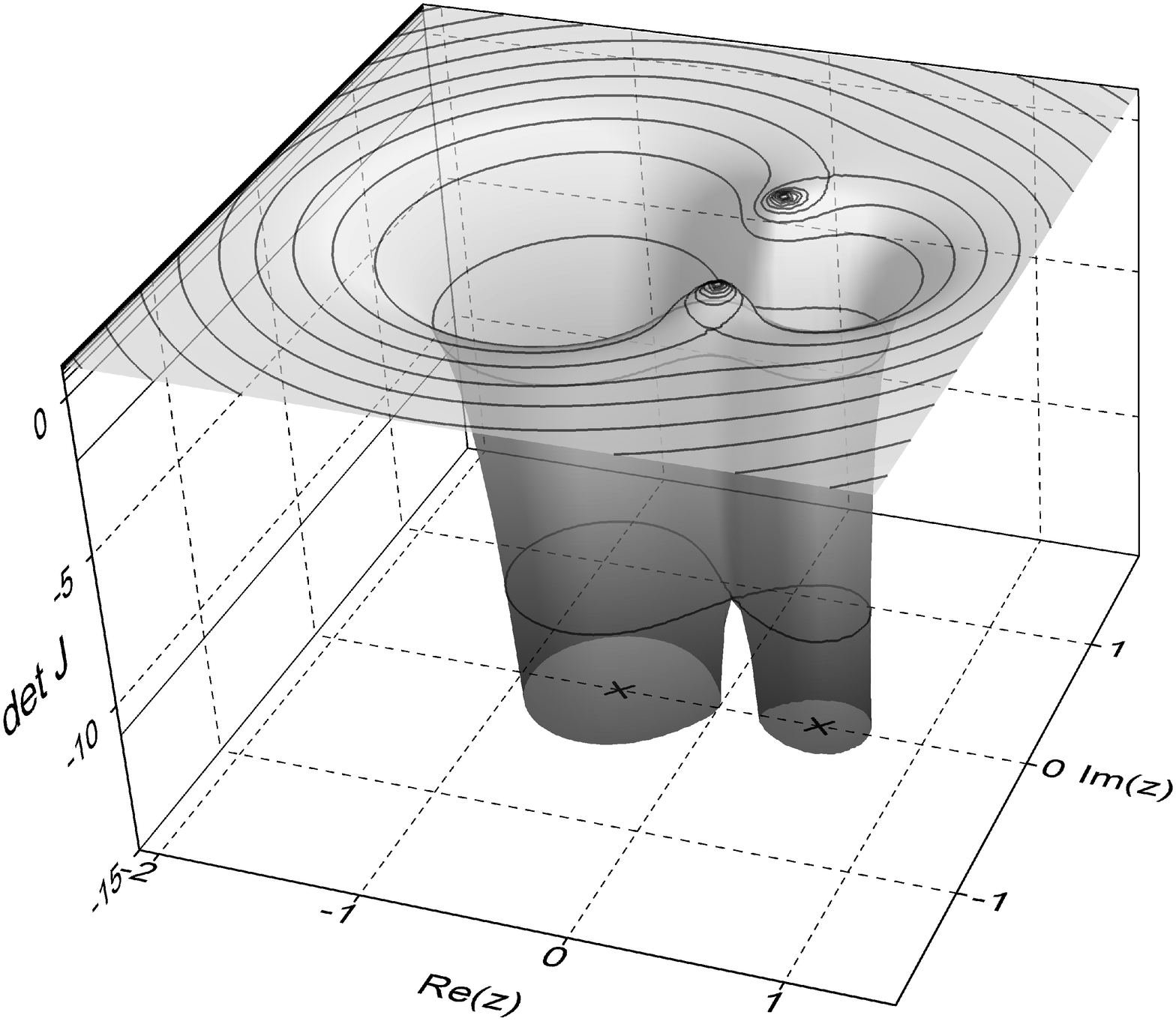}
\hspace{6mm}
\includegraphics[scale=.475]{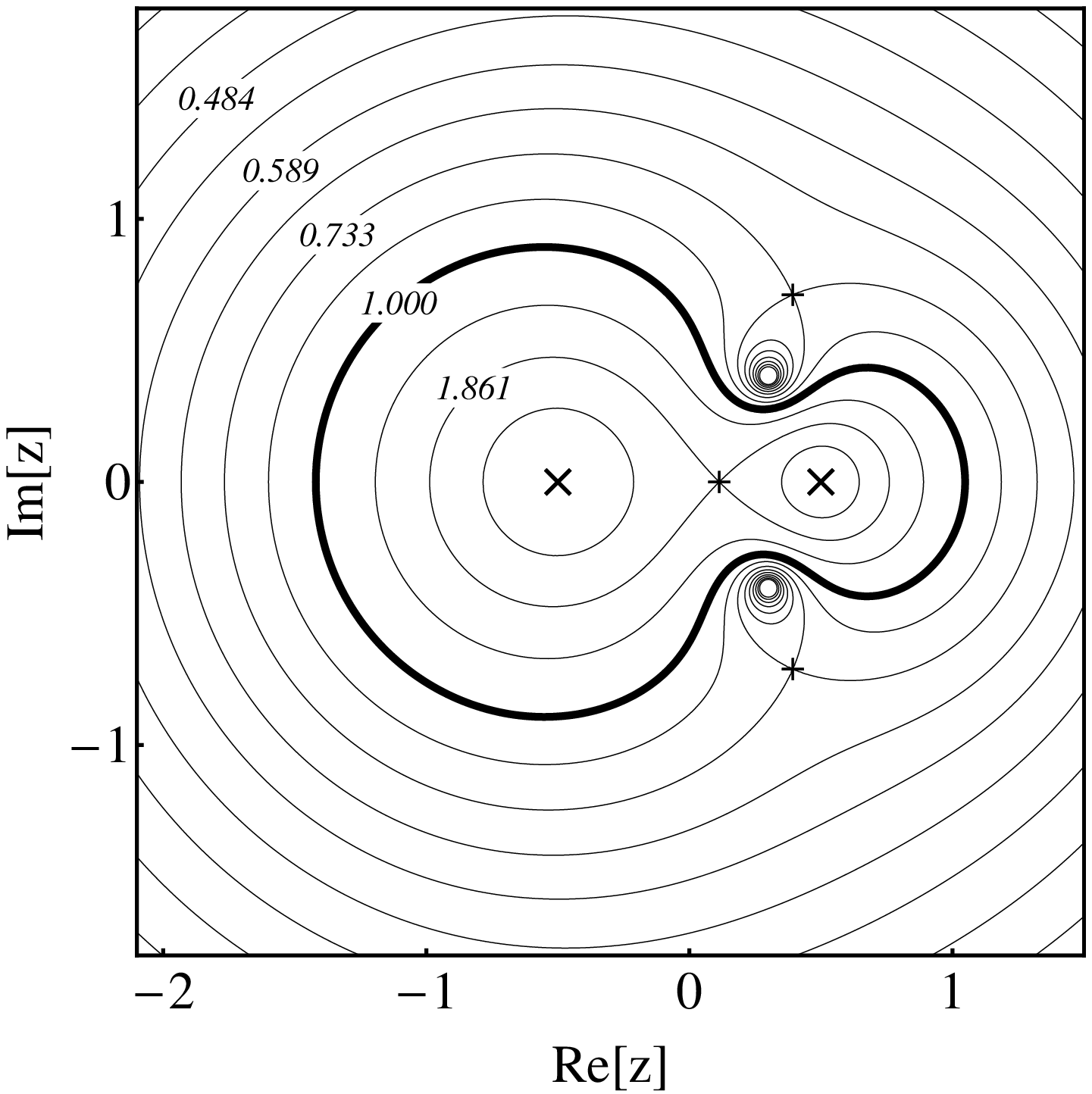}
\caption{Correspondence of Jacobian contours and critical curves of re-scaled lens configurations, illustrated on a binary lens with fractional masses $\{\mu_1,\mu_2\}=\{4/5,1/5\}$ and positions $\{z_1,z_2\}= \{-1/2,1/2\}$, i.e., separation $s_0=1$. Left panel: Jacobian surface of lens plotted over the image plane $z$, truncated below at ${\rm det}\,J=-15$, with contours plotted for ${\rm det}\,J\in\{{\bf -11.0},\, -2,\, {\bf 0},\, 0.5,\, {\bf 0.711},\, 0.82,\, {\bf 0.88},\, 0.92,\, {\bf 0.945},\, 0.962,\, 0.973\}$. Right panel: same contours with additional innermost ${\rm det}\,J=-100$ plotted in image plane, with the critical curve plotted in bold. Each contour is identically equal to the critical curve of a binary lens with separation and units on axes multiplied by $\sqrt[4]{1-{\rm det}\,J}$. Several separation values are included as labels for Jacobian contours listed in bold. Crosses: lens components; pluses: Jacobian saddle points. The two Jacobian maxima lie within the off-axis critical-curve loop sequences.}
\label{fig:Jacobian_correspondence}
\efi

\clearpage
\bfi
\hspace{-2mm}
\includegraphics[scale=.54]{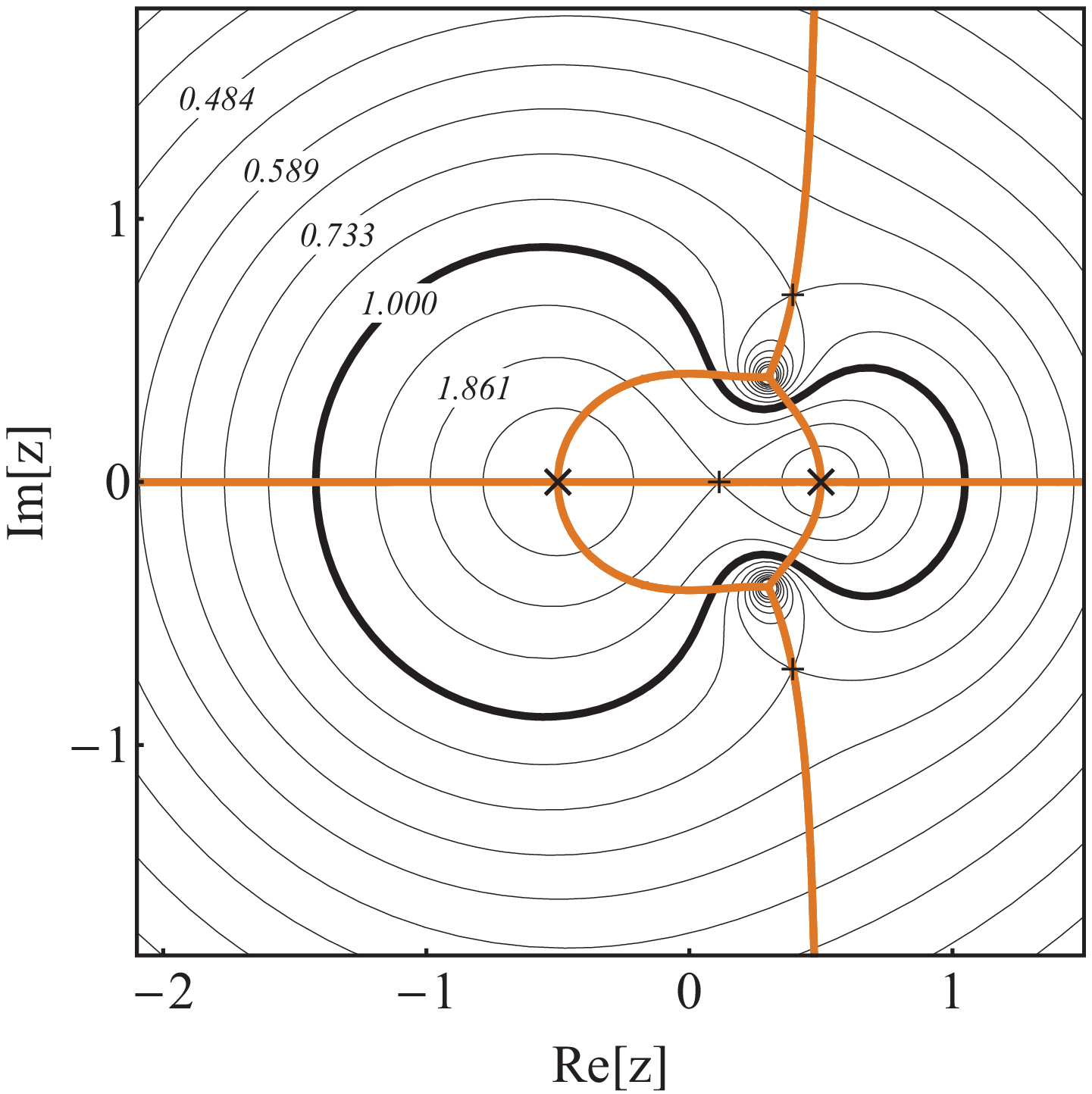}
\hspace{2mm}
\includegraphics[scale=.54]{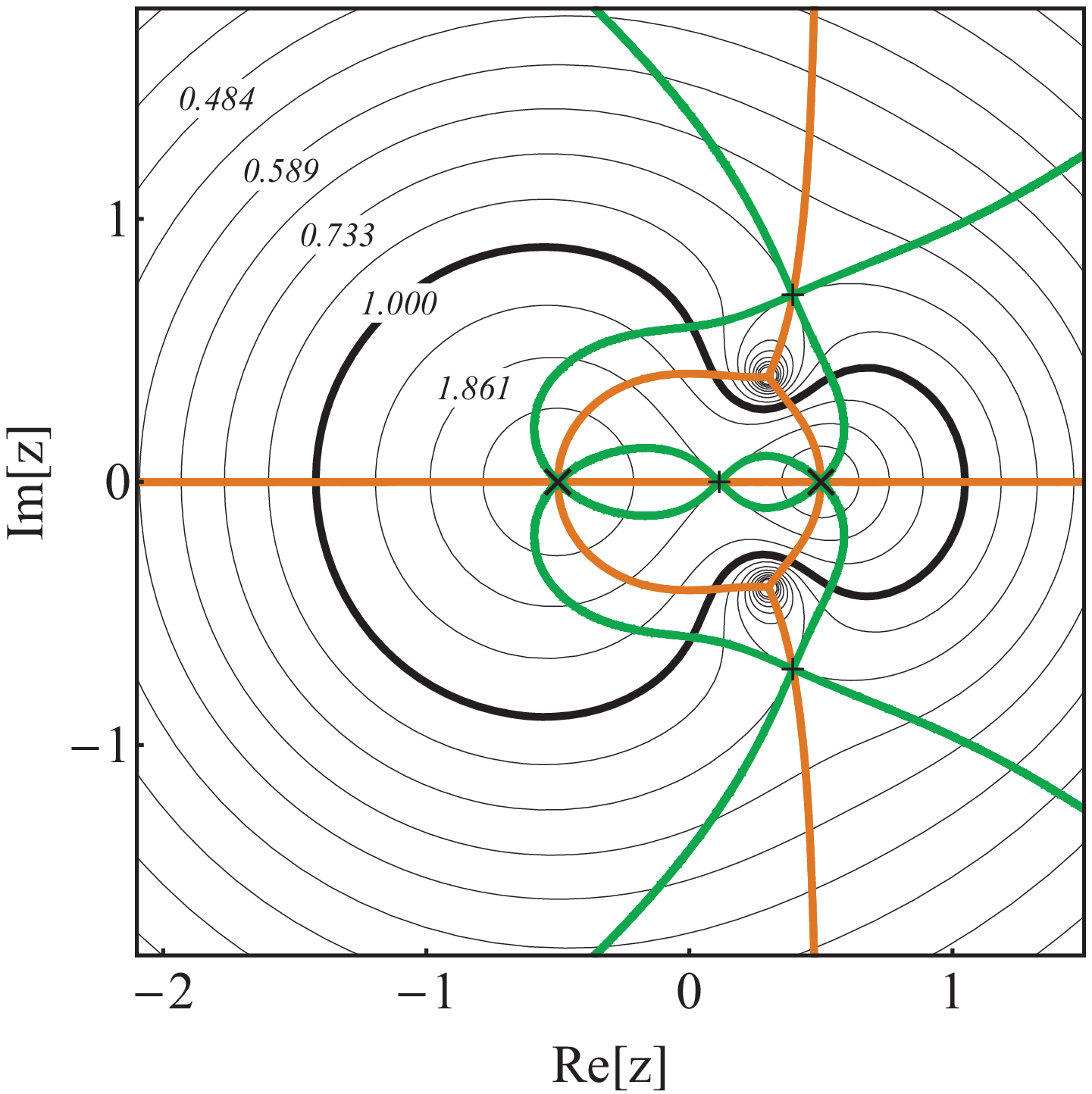}
\caption{Cusp curve (orange) and morph curve (green) of the binary lens from Figure~\ref{fig:Jacobian_correspondence}. The cusp curve intersects Jacobian contours at the positions of cusp images along the critical curves of the corresponding re-scaled lens configurations. The morph curve intersects the cusp curve at the lens positions and at the images of caustic-metamorphosis points for the corresponding re-scaled lens configurations. For a sequence of contours progressing through any such point the number of cusps on the corresponding caustics changes (see also Figure~\ref{fig:metamorphoses}). Here the intersections at Jacobian saddle points correspond to beak-to-beak metamorphoses. Notation as in Figure~\ref{fig:Jacobian_correspondence}.}
\label{fig:curves_binary}
\efi

\clearpage
\bfi
\hspace{-2mm}
\includegraphics[scale=.54]{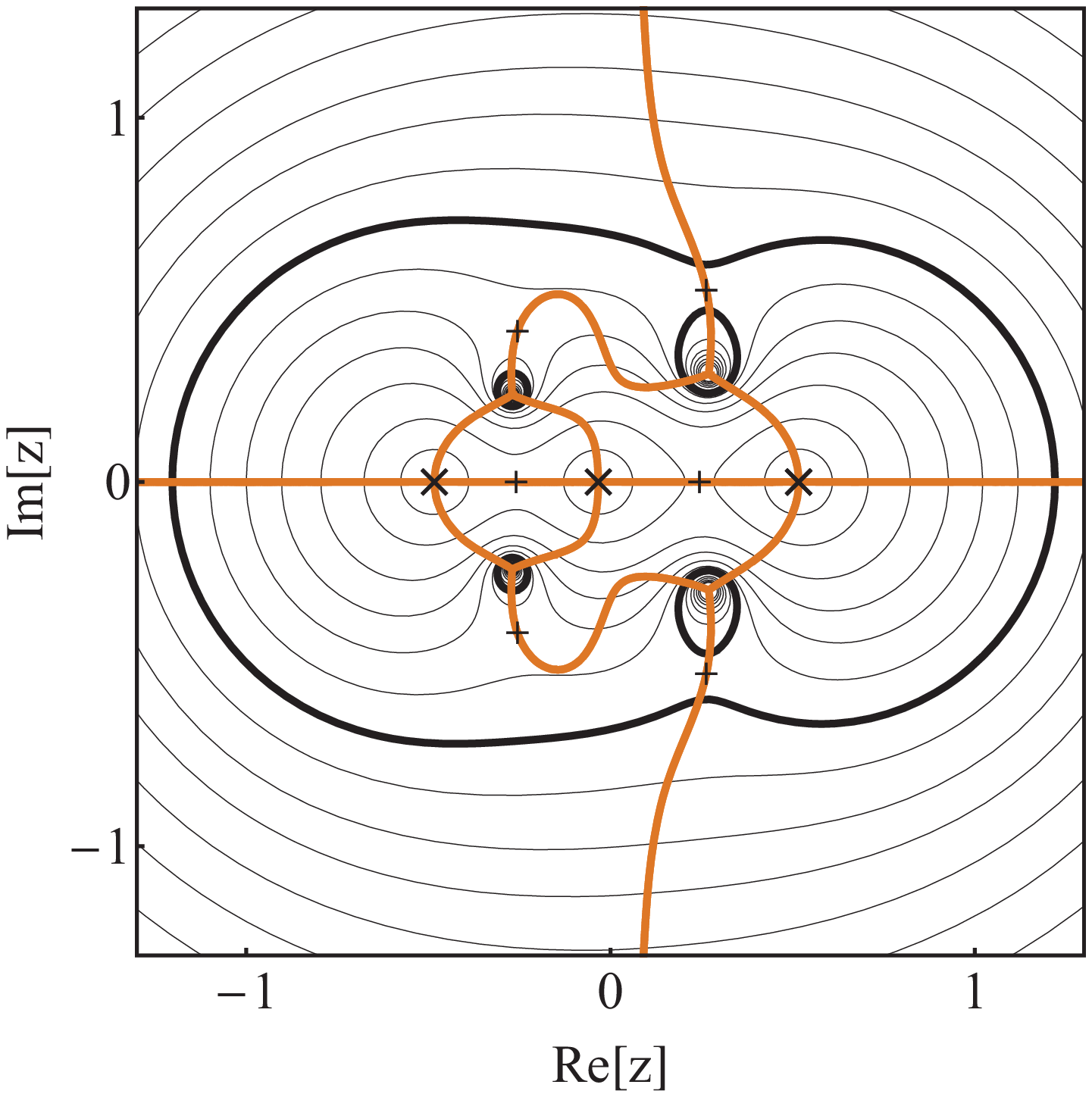}
\hspace{2mm}
\includegraphics[scale=.54]{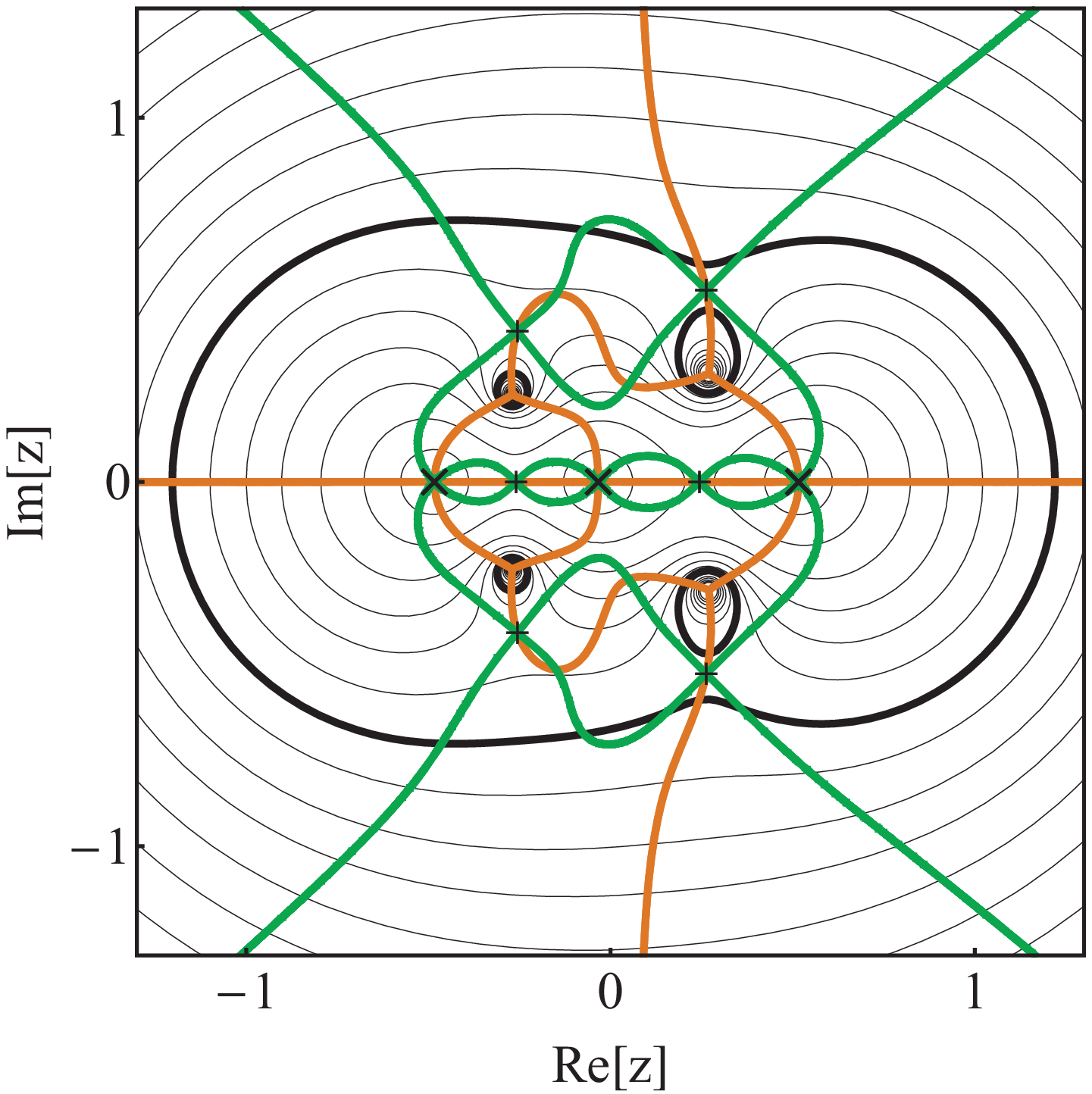}
\caption{Swallow-tail metamorphosis in the image plane: cusp curve (orange) and morph curve (green) of an equal-mass triple lens with positions $\{z_1,z_2,z_3\}= \{-29,-2,31\}\,s/30$, where $s=0.5$. Jacobian contours are plotted outward from the lenses for ${\rm det}\,J\in\{-1599,\,-99,\, -16.6,\,-5.25,\,-2.06,\,-0.69,\,0,\,0.36,\,0.578,\,0.698,\,0.788,\, 0.848,\,0.891\}$. The cusp-curve branch connecting the two Jacobian maxima above the real axis and its mirror counterpart below the real axis each pass through two points where they are tangent to the contours. These points, pinpointed in the right panel as simple intersections with the morph curve, correspond to swallow-tail metamorphoses. There are six additional saddle-point intersections corresponding to beak-to-beak metamorphoses. Notation as in Figure~\ref{fig:Jacobian_correspondence}.}
\label{fig:curves_triple1}
\efi

\clearpage
\bfi
\hspace{-2mm}
\includegraphics[scale=.54]{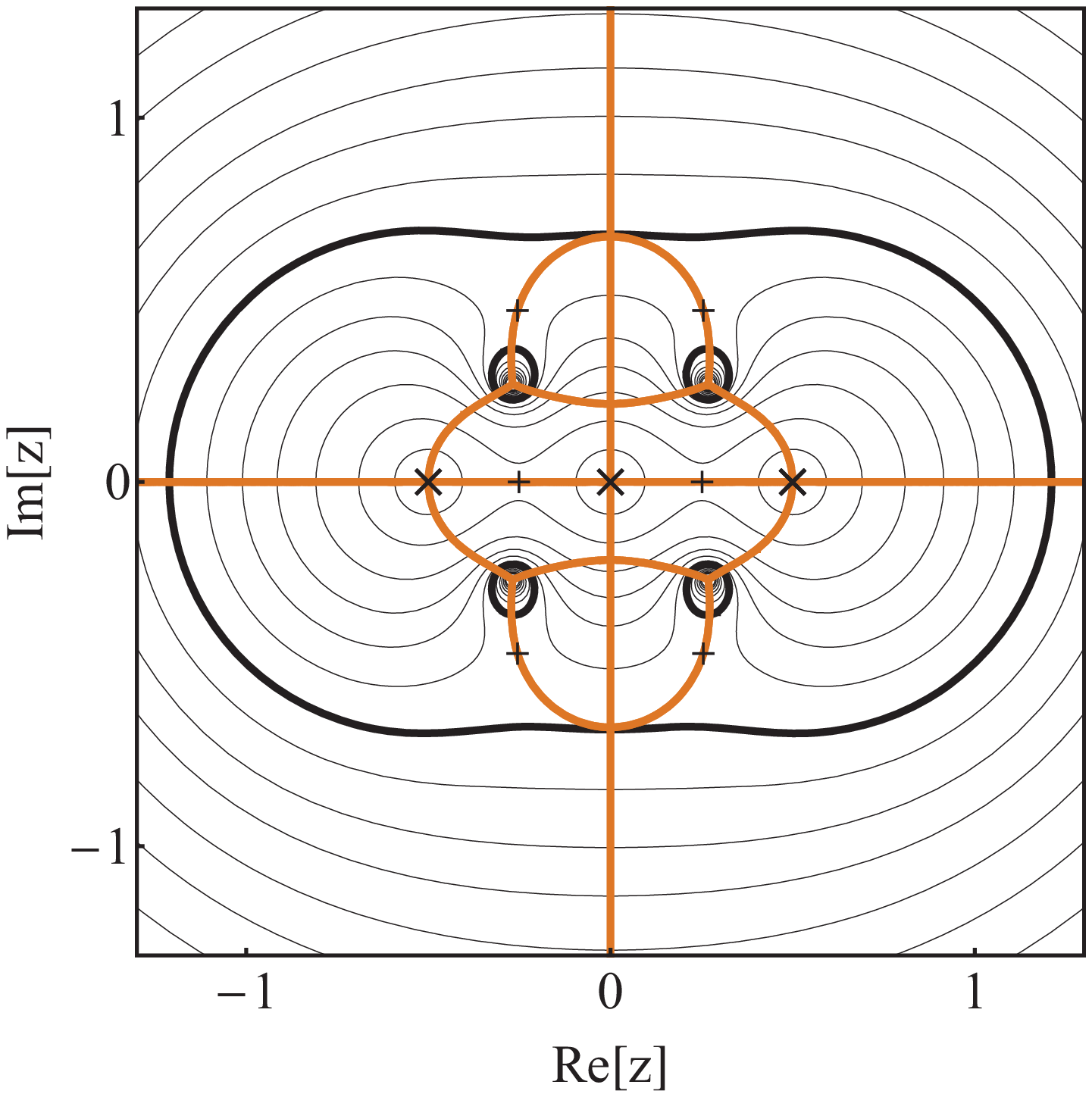}
\hspace{2mm}
\includegraphics[scale=.54]{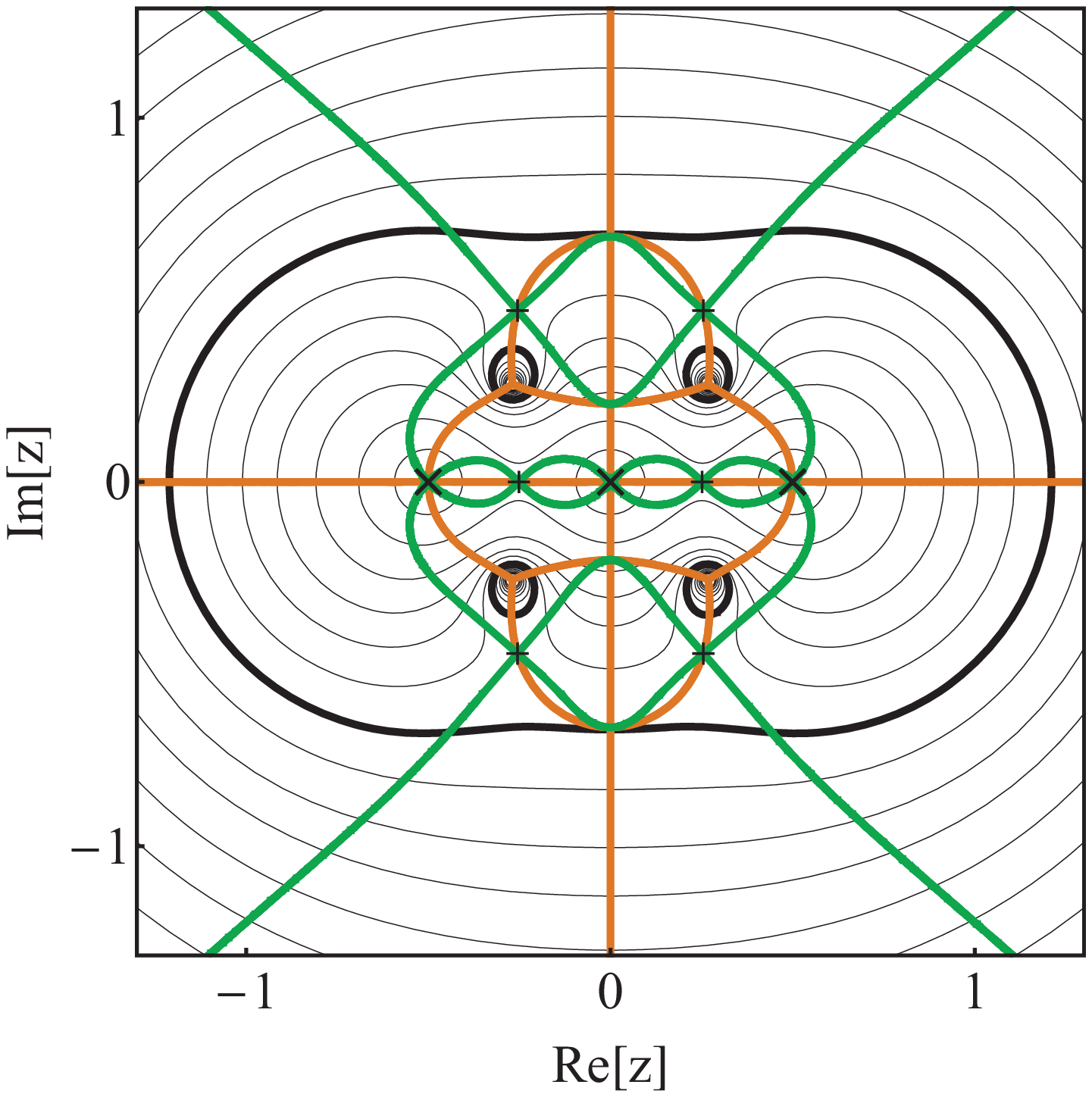}
\caption{Butterfly metamorphosis in the image plane: cusp curve (orange) and morph curve (green) of an equal-mass triple lens with positions $\{z_1,z_2,z_3\}= \{-1,0,1\}\,s$, where $s=0.5$. Jacobian contours are plotted outward from the lenses for ${\rm det}\,J\in\{-1599,\,-99,\,-16.6,\,-5.25,\, -2.06,\,-0.69,\,0,\,0.36,\,0.578,\,0.698,\,0.788,\,0.848,\,0.891\}$. The two cusp-curve branches connecting the Jacobian maxima above the real axis and their mirror counterparts below the real axis each pass through a self-intersection point at which they are tangent to the contours. These points, found in the right panel as tangent intersections with the morph curve, correspond to butterfly metamorphoses. There are six additional beak-to-beak metamorphoses at saddle-point intersections. Notation as in Figure~\ref{fig:Jacobian_correspondence}.}
\label{fig:curves_triple2}
\efi

\clearpage
\bfi
\includegraphics[width=16.5cm]{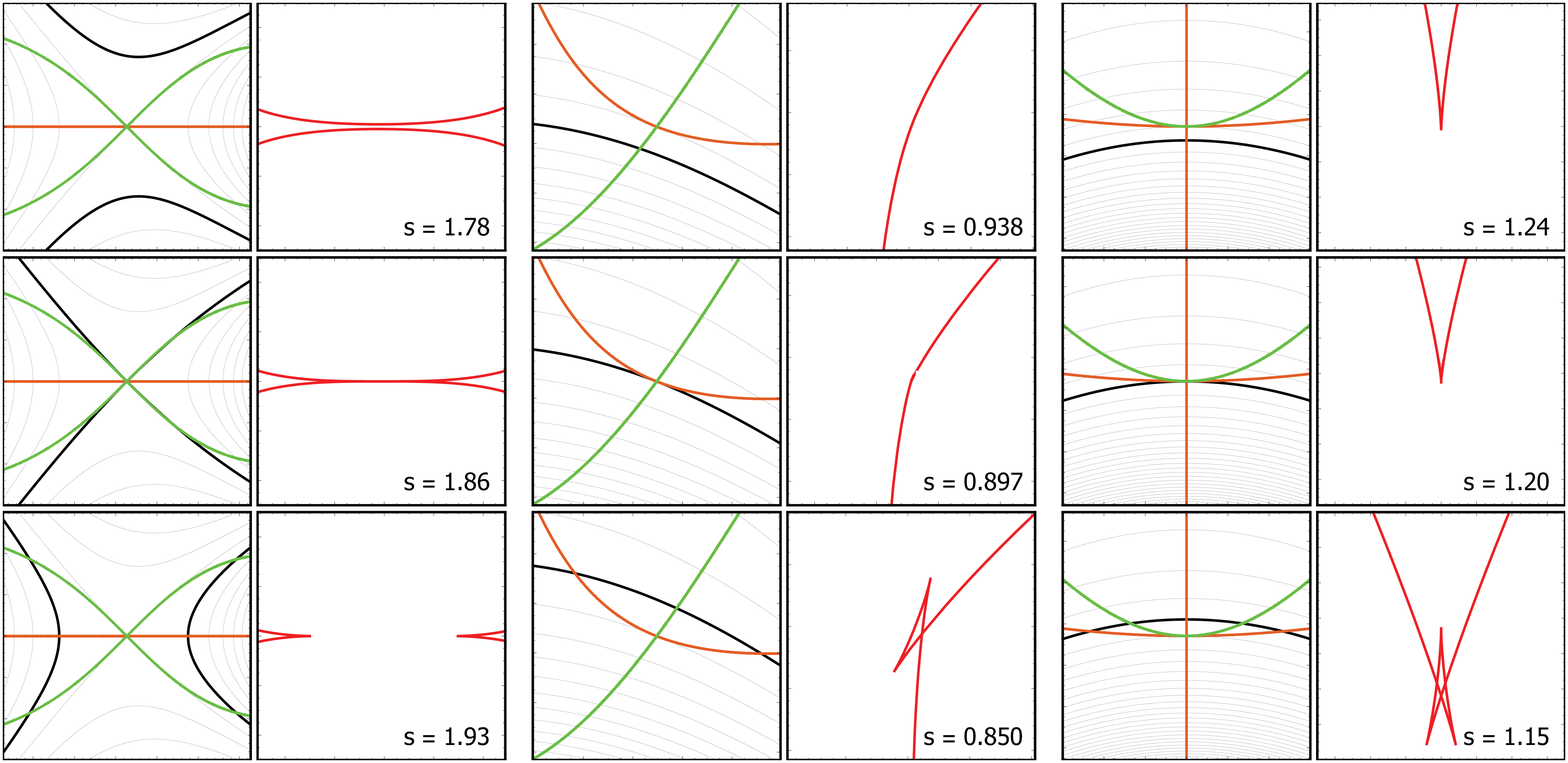}
\caption{First three elementary caustic metamorphoses of the $n$-point-mass lens. Left column: beak-to-beak (detail from Figure~\ref{fig:curves_binary}); central column: swallow-tail (detail from Figure~\ref{fig:curves_triple1}); right column: butterfly metamorphosis (detail from Figure~\ref{fig:curves_triple2}). Left panel in each pair: detail of image plane divided by scale parameter $s$ (black: critical curve; gray: other Jacobian contours; orange: cusp curve; green: morph curve). Right panel in each pair: detail of source plane (red: caustic). Rows correspond to scale parameters $s$ marked in the panels. Top: before metamorphosis with $n$ cusps; middle: at metamorphosis; bottom: after metamorphosis with $n+2$ cusps.}
\label{fig:metamorphoses}
\efi

\end{document}